%% file: main.tex
\documentclass[sigconf,10pt]{acmart}
\acmYear{2025}\copyrightyear{2025}
\setcopyright{acmlicensed}
\acmConference[SOSP '25]{ACM SIGOPS 31st Symposium on Operating Systems Principles}{October 13--16, 2025}{Seoul, Republic of Korea}
\acmBooktitle{ACM SIGOPS 31st Symposium on Operating Systems Principles (SOSP '25), October 13--16, 2025, Seoul, Republic of Korea}
\acmDOI{10.1145/3731569.3764849}
\acmISBN{979-8-4007-1870-0/25/10}
\usepackage{amsmath}
\usepackage{amsthm}

\usepackage{amsfonts}
\usepackage{mathtools}
\usepackage{ifthen}
\usepackage{graphicx}
\usepackage{caption}
\usepackage{subcaption}
\usepackage[T1]{fontenc}
\usepackage[utf8]{inputenc}
\usepackage[frozencache,cachedir=.]{minted}
\usepackage[most,minted]{tcolorbox}
\PassOptionsToPackage{dvipsnames}{xcolor}
\definecolor{Blue}{RGB}{68,114,196}
\definecolor{LightBlue}{RGB}{218,227,243}
\definecolor{VeryLightBlue}{RGB}{238,241,255}
\definecolor{LightGreen}{RGB}{226,240,217}
\definecolor{Pink}{RGB}{245,210,248}
\definecolor{LightOrange}{RGB}{251,229,214}
\definecolor{Orange}{RGB}{237,125,49}
\definecolor{LightGray}{gray}{0.9}
\definecolor{VeryLightGray}{gray}{0.95}
\newlength{\charlen}
\tcbset{on line,
        boxsep=2pt, left=\charlen,right=0pt,top=\charlen,bottom=0pt,
        colframe=white,colback=VeryLightGray,
        titlerule=0.0pt,
        highlight math style={enhanced}
        }
\newtcbox{\LightBlueBox}{colback=LightBlue}
\newtcbox{\LightGreenBox}{colback=LightGreen}
\newtcbox{\PinkBox}{colback=Pink}
\newtcbox{\LightOrangeBox}{colback=LightOrange}
\usepackage{pifont}

\usepackage{tabularx}

\usepackage{multirow}
\usepackage{makecell}
\setcellgapes{0pt}
\newcolumntype{Y}{>{\centering\arraybackslash}X}
\setminted[py]{escapeinside=||,mathescape=true,fontfamily=zi4}
\usepackage{listings}

\newtcblisting[auto counter]{CodeListing}[3][]{
enhanced,
listing engine=minted,
minted language=py,
listing only,
minted options = {
    linenos, 
    numbersep=2mm,
    #3
},
overlay={%
    \begin{tcbclipinterior}
        \fill[gray!25] (frame.south west) rectangle ([xshift=4mm]frame.north west);
    \end{tcbclipinterior}
},
fonttitle=\fontfamily{\sfdefault}\selectfont,
colback=VeryLightGray,left=4mm,%right=0pt,top=0pt,bottom=0pt,
colbacktitle=LightBlue,coltitle=black,
title={Listing \thetcbcounter: #1},
#2
}
\newtcolorbox{CodeListingInSubfigure}[1][]{
enhanced,breakable,
fonttitle=\fontfamily{\sfdefault}\selectfont,
left=1pt,right=0pt,top=0pt,bottom=0pt,
#1
}

\DeclareRobustCommand\SystemName{DCP}
\DeclareRobustCommand\cd{attention}
\DeclareRobustCommand\ci{context-independent}

\begin{document}

\title[]{\SystemName{}: Addressing Input Dynamism In Long-Context Training via Dynamic Context Parallelism}
\author{Chenyu Jiang}
\authornote{Work done during internship at AWS.}
\email{jchenyu@connect.hku.hk}
\orcid{0009-0006-5714-3872}
\affiliation{%
  \institution{The University of Hong Kong}
  \country{}
}

\author{Zhenkun Cai}
\email{zkcai@amazon.com}
\affiliation{%
  \institution{Amazon Web Services}
  \country{}
}

\author{Ye Tian}
\email{yetiansh@connect.hku.hk}
\authornotemark[1]
\affiliation{%
  \institution{The University of Hong Kong}
  \country{}
}

\author{Zhen Jia}
\email{zhej@amazon.com}
\affiliation{%
  \institution{Amazon Web Services}
  \country{}
}

\author{Yida Wang}
\email{wangyida@amazon.com}
\affiliation{%
  \institution{Amazon Web Services}
  \country{}
}

\author{Chuan Wu}
\email{cwu@cs.hku.hk}
\affiliation{%
  \institution{The University of Hong Kong}
  \country{}
}


\begin{abstract}
Context parallelism has emerged as a key technique to support long-context training, a growing trend in generative AI for modern large models. However, existing context parallel methods rely on static parallelization configurations that overlook the dynamic nature of training data, specifically, the variability in sequence lengths and token relationships (i.e., attention patterns) across samples. As a result, these methods often suffer from unnecessary communication overhead and imbalanced computation.
In this paper, we present DCP, a dynamic context parallel training framework that introduces fine-grained blockwise partitioning of both data and computation. By enabling flexible mapping of data and computation blocks to devices, DCP can adapt to varying sequence characteristics, effectively reducing communication and improving memory and computation balance.
Micro-benchmarks demonstrate that DCP accelerates attention by 1.19x\textasciitilde{}2.45x under causal masks and 2.15x\textasciitilde{}3.77x under sparse attention patterns.
Additionally, we observe up to 0.94x\textasciitilde{}1.16x end-to-end training speed-up for causal masks, and 1.00x\textasciitilde{}1.46x for sparse masks.
\end{abstract}





% \maketitle

\maketitle

%-------------------------------------------------------------------------------
\input{sections/1_introduction}
%-------------------------------------------------------------------------------
\input{sections/2_background_motivation}
%-------------------------------------------------------------------------------
\input{sections/3_system}
%-------------------------------------------------------------------------------
\input{sections/4_dynamic_context_parallel}
%-------------------------------------------------------------------------------
\input{sections/5_executor}
%-------------------------------------------------------------------------------
\input{sections/6_other_impl_details}
%-------------------------------------------------------------------------------
\input{sections/7_evaluation}
%-------------------------------------------------------------------------------
\input{sections/8_related_work}
%-------------------------------------------------------------------------------

\begin{acks}
We would like to thank the anonymous reviewers and our shepherd Tim Harris for their valuable feedback.
This work was supported in part by grants from Hong Kong RGC under the contracts 17204423, 17205824, and C7004-22G (CRF).
\end{acks}

\newpage
\bibliographystyle{plain}
\bibliography{bibliography}

\end{document}

%% file: sections/1_introduction.tex
\section{Introduction}

State-of-the-art AI systems have achieved remarkable performance across a diverse range of tasks~\cite{openai2024gpt4ocard,deepseekai2025deepseekr1,anthropic2025claude3,google2025gemini,intelligence2024amazon}.
A notable trend in modern large models is the increasing context length (number of tokens as input), meant for enhancing deep learning models' capacity in processing extensive amounts of information (e.g., long documents and code-bases).
For example, GPT-4o supports a 128K context window~\cite{openai2024gpt4ocard};
Claude 3.5 Sonnet extends the context window size to 200K~\cite{anthropic2025claude3};
Gemini 2.5 Pro scales the context window to 2M tokens~\cite{google2025gemini}.
The increased context length greatly raises the memory and computation requirements of state-of-the-art large generative models, making them significantly more expensive to train. 

\begin{figure}
    \centering
    \includegraphics[width=\linewidth]{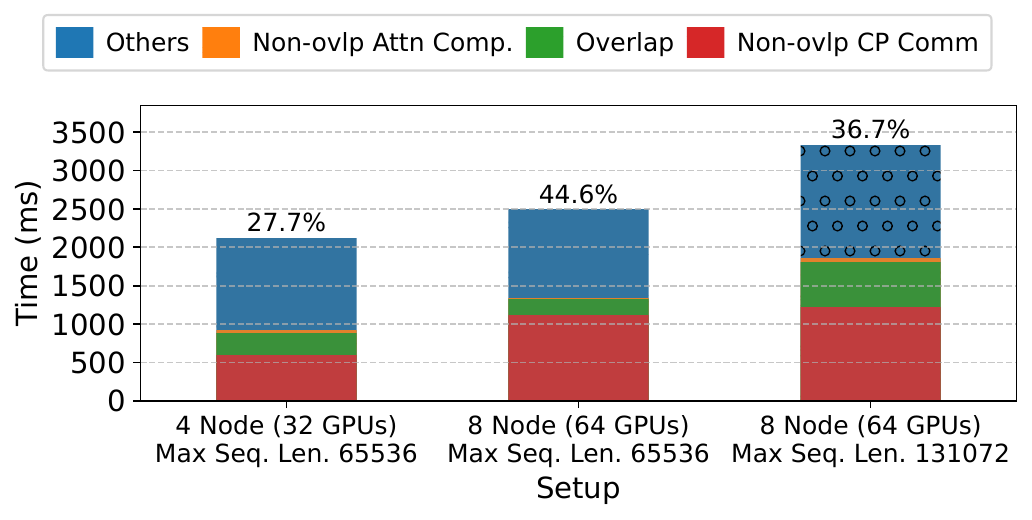}
    \vspace{-8mm}
    \caption{CP communication overhead when training a 8B GPT model on an Amazon EC2 \texttt{p4d.24xlarge} cluster (400Gbps interconnect between nodes) with 4-way tensor and 16-way context parallelism, using the LongAlign~\cite{bai2024longalign} dataset. Overlap: overlapping CP communication and attention computation. Communication overhead fraction (vs. total iteration time) is shown above each bar.}
    \label{fig:iter_time_decomposition}
    \vspace{-8mm}
\end{figure}

To address this challenge, recent approaches adopt context parallelism (CP), which partitions each sequence in the training data evenly across all devices~\cite{jacobs2023deepspeedulysses,liu2024ringattention,fang2024usp,gu2024loongtrain}.
These methods reduce memory consumption and enable training with longer context lengths, but incur additional communication overhead~\cite{dubey2024llama3herdmodels,xue2024longvila}.
Notably, this communication cost increases with the size of the training cluster (Fig.~\ref{fig:iter_time_decomposition}).
As both model sizes and context lengths continue to grow, the communication overhead is expected to increase significantly.

Existing CP approaches uniformly apply a fixed parallelization configuration (data partitioning and placement) for all batches.
Such static partitioning methods fail to account for the inherent dynamism in training data, which we categorize into:
1. the variance in input sequence lengths, and
2. the variance in token relationships within each sequence.
As a result, these methods miss key opportunities for optimization.

\begin{figure}
    \centering
    \includegraphics[width=0.8\linewidth]{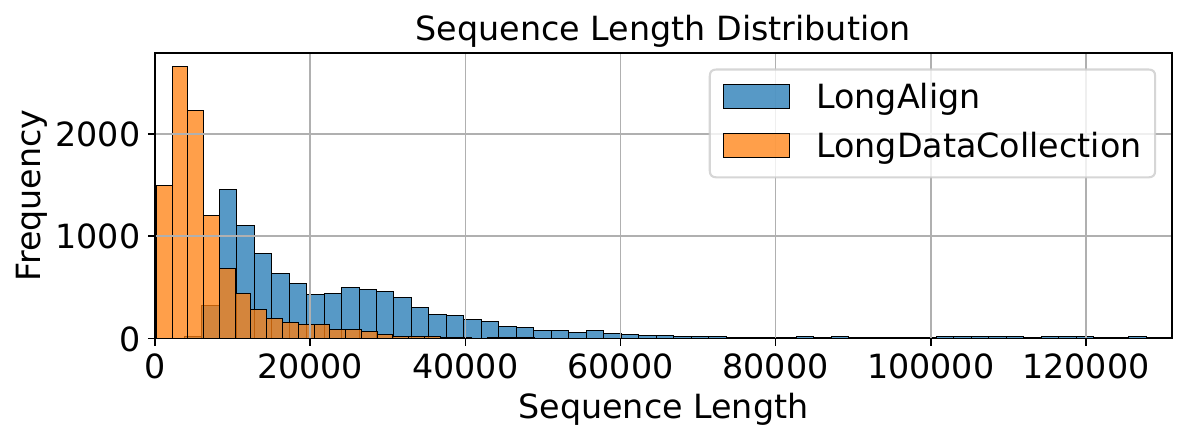}
    \vspace{-4mm}
    \caption{Sequence length distribution of LongAlign~\cite{bai2024longalign} and LongDataCollection~\cite{togethercomputer2024longdatacollections} datasets, capped at 131072.}
    \label{fig:dataset_seqlen_distribution}
    \vspace{-8mm}
 \end{figure}

\textbf{Variance in input sequence lengths.}
Modern training datasets often exhibit a highly skewed distribution of sequence lengths, especially in long-context settings, where shorter sequences are significantly more common than longer ones~\cite{jiang2024dynapipe, ge2025bytescale}.
For instance, during the supervised fine-tuning phase of Llama 3 training, long-context samples constitute only 0.11\% of the dataset~\cite{dubey2024llama3herdmodels}.
Similar patterns can be observed in other datasets (see Fig.~\ref{fig:dataset_seqlen_distribution}).
Larger pre-training datasets, such as The Pile~\cite{gao2020thepile}, also exhibit similar document length distributions.
Static parallelization can introduce redundant communication when processing shorter sequences, thereby increasing overall execution time.

\textbf{Variance in token relationships (attention patterns)}.
In modern LLMs, token relationships are typically expressed through attention masks.
Existing static context parallelization schemes are primarily designed for causal attention~\cite{liu2024ringattention,fang2024usp,gu2024loongtrain}.
However, recent studies have advocated the use of diverse attention masks to accelerate training or address novel training scenarios.
For example, in reinforcement learning-based post-training, a shared question mask can eliminate redundant computation between a question and its multiple answers~\cite{wang2025flashmask}. 
Sliding-window or lambda-shaped masks~\cite{han2024lminfinite, lin2024infinitellm} are widely used to significantly sparsify attention and reduce required computation.
These sparse or structured attention masks break the assumptions on attention workload distribution that static methods rely on.
As a result, applying static partitioning in such settings leads to severe load imbalance and redundant communication, which undermines performance.

To address the limitations of static parallelization configurations in current context parallelism approaches, we propose a dynamic context parallelism approach that constructs a different parallelization configuration for each training iteration.
We systematically model context-parallel training by partitioning attention inputs and outputs into fine-grained data blocks and constructing computation blocks that capture attention patterns.
These blocks can be flexibly assigned to different devices, enabling customized parallelism and data/computation placement configurations tailored to each sequence.
We optimize the placement of data and computation blocks in each iteration by formulating it as a hypergraph partitioning problem, aiming to minimize communication costs while satisfying memory and compute balance constraints.
We also automatically generate computation and communication schedules for the blocks assigned to each device, forming pipelines to overlap their execution.
This dynamic planning is managed by a data loader wrapper, which pre-fetches data and serializes device-specific schedules using five distinct instructions prior to the corresponding iteration.
A custom executor efficiently executes these instructions using fused kernels, minimizing the overhead associated with fine-grained parallelism.

We summarize our contributions as follows:

$\triangleright$ We devise a representation for both data and computation that explicitly captures the effect of input dynamism in context parallelism, 
allowing us to systematically define fine-grained parallelization configurations for each training iteration.

$\triangleright$ We formulate the problem of optimizing the parallelization configuration with hypergraph partitioning, enabling efficient solutions using established algorithms~\cite{schlag20kahypar}.

$\triangleright$ We provide an end-to-end framework implementation (i.e., DCP) that enables long context training with dynamic context parallelism, minimizing planning and runtime overheads.

$\triangleright$ We perform an extensive evaluation against state-of-the-art CP frameworks including TransformerEngine~\cite{nvidia2024transformerengine} and LoongTrain~\cite{gu2024loongtrain}.
Micro-benchmarks show that DCP achieves 1.19x\textasciitilde{}2.45x speed-up with causal, and 2.15x\textasciitilde{}3.77x with sparse attention masks for individual attention layers.
End-to-end experiments show a 0.94x\textasciitilde{}1.16x speed-up with causal mask, and 1.00x\textasciitilde{}1.46x with sparse attention masks.

% \vspace{5mm}

%% file: sections/2_background_motivation.tex
\section{Background and Motivation}

\subsection{Block-wise attention computation}
\begin{figure}
     \centering
     \includegraphics[width=0.9\linewidth]{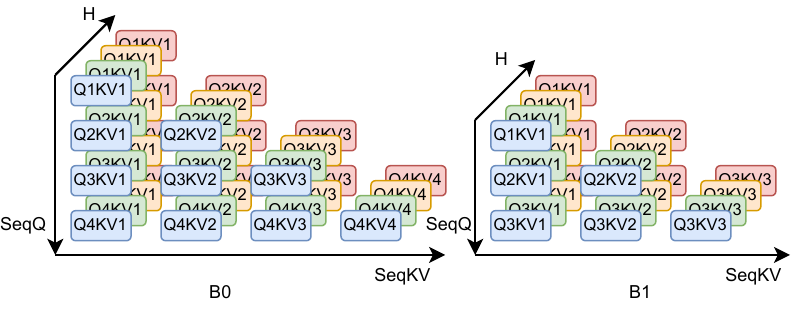}
     \caption{Four parallelizable dimensions of an attention operator.
     Each block represents the computation of $\mathbf{\hat{O}_{bhqk}}$ (Listing~\ref{lst:parallel_attn_algo} line 5) with corresponding Q and KV blocks. 
     Figure shows attention of two sequences (B0 and B1) with lengths 4 and 3 tokens, respectively, using causal mask.}
     \label{fig:attn_parallel_dims}
\end{figure}

Attention is one of the central components in transformer-based large models~\cite{vaswani2017attention}.
The standard masked self-attention is:
\[\mathbf{O}_{bh::}=\text{RowWiseSoftmax}(\frac{(\mathbf{Q}_{bh::} \mathbf{K}_{bh::}^T) \odot \mathbf{M}_{b::}}{\sqrt{D}})\mathbf{V}_{bh::}\]
\noindent where $\mathbf{Q}$ (query), $\mathbf{K}$ (key), $\mathbf{V}$ (value), $\mathbf{O}$ (attention output) are 4-dimensional tensors of shape $[B, H, L, D]$. $B$ is the input batch size, $H$ is the number of attention heads, $L$ is the input sequence length and $D$ is head dimension size.
Subscripts $bh\mathpunct{::}$ are indices into the respective dimensions, i.e., $\mathbf{Q}_{bh::}$ is a slice (matrix of shape $[L, D]$) of the tensor $\mathbf{Q}$ at index $b$ in the first dimension ($B$) and $h$ in the second dimension ($H$).
$\mathbf{M}$ is a boolean mask of shape [B, L, L], zeroing out unwanted interactions between tokens during attention computation (i.e., the attention mask).
Using the online softmax trick~\cite{dao2022flashattention, rabe2022selfattentiondoesneedon2}, attention can be computed block-wise, with each block processed in parallel.
Suppose that we divide each tensor into $\mathcal{B}_{b}$ blocks along the batch dimension and $\mathcal{B}_{h}$ blocks along the head dimension;
then, along the sequence length dimension, we divide $\mathbf{Q}$ into $\mathcal{B}_{q}$ blocks, and $\mathbf{K,V}$ into $\mathcal{B}_{kv}$ blocks, the parallel attention computation can be expressed in pseudo code as:

\noindent
\begin{CodeListing}[Block parallel attention]{label=lst:parallel_attn_algo}{}
parallel for b in [1, |$\mathcal{B}_b$|]:
  parallel for h in [1, |$\mathcal{B}_h$|]:
    parallel for q in [1, |$\mathcal{B}_q$|]:
      parallel for k in [1, |$\mathcal{B}_{kv}$|]:
        |$\mathbf{\hat{O}_{bhqk}}$| = Softmax(|$\frac{(\mathbf{Q_{bhq:}K^T_{bhk:}})\odot \mathbf{M_{bqk}}}{\sqrt{D}})\mathbf{V_{bhk:}}$|
      |$\mathbf{O_{bhq:}}$| = RescaleAndSum(|$\mathbf{\hat{O}}_{bhq1},\ldots,\mathbf{\hat{O}}_{bhq\mathcal{B}_{kv}}$|)
\end{CodeListing}

\noindent
where the subscripts are now block indices.
Such block-wise parallelization is widely adopted by efficient GPU attention kernel libraries like FlashAttention~\cite{dao2022flashattention}, as it eliminates the memory access cost of materializing the large intermediate tensor $\mathbf{QK^T}$.
With appropriate block sizes, Q or KV blocks can also reside in the fast GPU shared memory, enabling efficient reuse and further reducing memory access costs.
Therefore, we also use such block division design to analyze distributed attention and identify the following four parallelizable dimensions for the attention operator: \textbf{batch}, \textbf{head}, \textbf{SeqQ} and \textbf{SeqKV}, corresponding to the first four lines in Listing~\ref{lst:parallel_attn_algo}, respectively (Fig.~\ref{fig:attn_parallel_dims}).
$\mathbf{\hat{O}_{bhqk}}$ (Line 5) with different subscripts can potentially be computed in parallel across devices.
Since $\mathbf{K}$ and $\mathbf{V}$ are always partitioned and parallelized together, we refer to them as a whole as $\mathbf{KV}$, in the rest of the paper.

\subsection{Context parallelism}
\label{sec:background_cp}

\begin{figure}
    \centering
    \includegraphics[width=\linewidth]{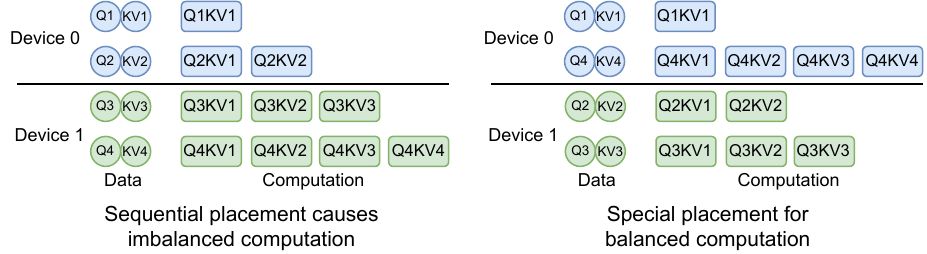}
    \caption{Special data and computation placement for causal mask.}
    \label{fig:causal_placement}
\end{figure}

Operators in a transformer-based large model can be categorized into: 1) attention operators, whose computation of tokens depends on other tokens in an input sequence; 2) context-independent operators, in which computations of tokens are independent (e.g., layer normalization and MLP layers).
Context parallelism of an operator parallelizes the computation of tokens within each input sequence among devices.
Context parallelizing the attention operator must be handled carefully, as memory/computation balance and communication overhead are all critical to performance. 
Context parallelizing context-independent operators is more straightforward, as it does not involve additional communication overhead during parallel token processing by different devices.
Context parallelism's token-to-device assignment is usually inherently decided by input and output data (token) placement of connected attention layers.

Existing context parallelism performs distributed attention at the head and SeqQ dimensions, with fixed communication schedules as well as input data and computation placements.
When parallelizing at the SeqQ dimension, each device is responsible for the computation of certain tokens in a query with all KVs, with the communication for KV forming a ring pattern~\cite{liu2024ringattention,zhu2024ringflashattn}.
When parallelizing at the head dimension, each device calculates different attention heads and needs to access Q, KV and output of assigned heads of all sequences, resulting in all-to-all communications~\cite{jacobs2023deepspeedulysses}.
The two parallelization schemes can also be applied jointly~\cite{fang2024usp,gu2024loongtrain,nvidia2024transformerengine}.
Most methods~\cite{zhu2024ringflashattn,fang2024usp,gu2024loongtrain,nvidia2024transformerengine} use a placement that optimizes computation balance under the causal mask:
With $R$ devices, each input sequence is uniformly sliced into $2R$ chunks, and the $i$th device is assigned the $i$th chunk and the $(2R-i+1)$th chunk (Fig.~\ref{fig:causal_placement}).

\subsection{Redundant communication on variable-length inputs}
\label{sec:moti_seqlen_distribution}

\begin{figure}
     \centering
     \begin{subfigure}[t]{\linewidth}
         \centering
         \includegraphics[width=\linewidth]{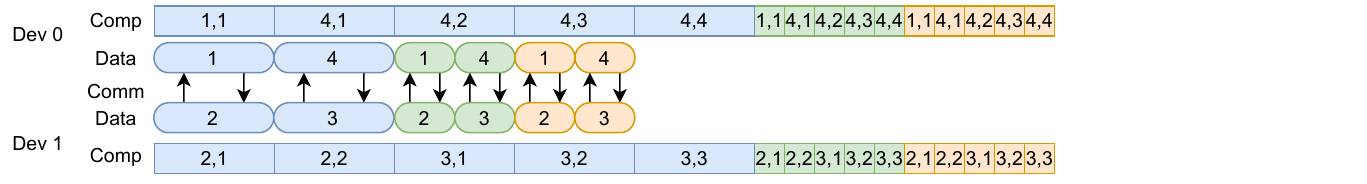}
         \caption{Existing CP designs: heavy communication is required.}
         \label{fig:moti_exp_current_cp}
     \end{subfigure}
     \hfill
     \begin{subfigure}[t]{\linewidth}
         \centering
         \includegraphics[width=\linewidth]{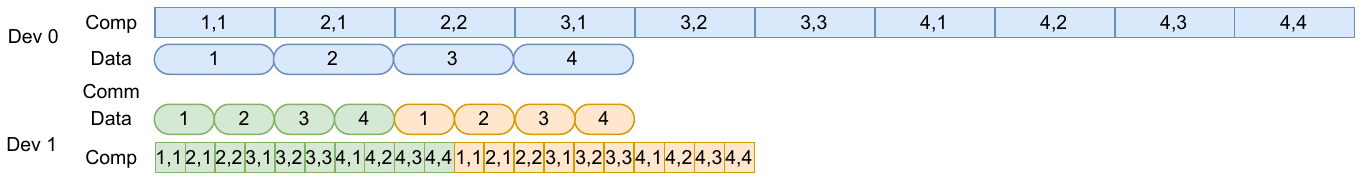}
         \caption{DP: no communication, but imbalanced computation.}
         \label{fig:moti_exp_dp}
     \end{subfigure}
     \hfill
      \begin{subfigure}[t]{\linewidth}
         \centering
         \includegraphics[width=\linewidth]{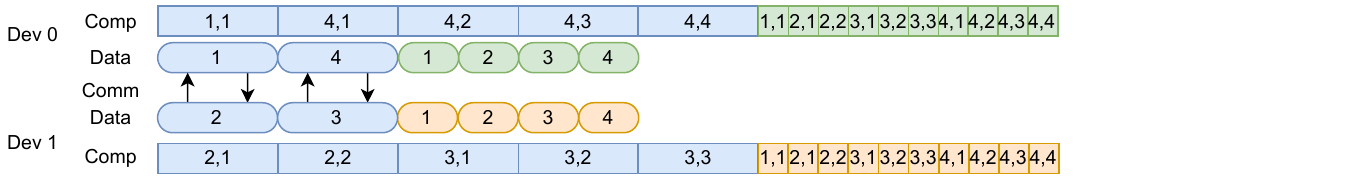}
         \caption{CP applied to the long sequence and DP applied between short sequences: balanced computation, less communication.}
         \label{fig:moti_exp_optimal_placement}
     \end{subfigure}
     \caption{Communication and computation required under different parallelization configurations.
        Color distinguishes different sequences.
        Round-cornered blocks: input token blocks (4 blocks in each sequence), where each block in the blue sequence contains twice as many tokens as green and orange sequences.
        Square blocks $[a, b]$: attention between token block $a$ and token block $b$. Arrow: communication of KV for corresponding blocks.}
        \label{fig:sub_optimal_ringattn}
\end{figure}

We observe communication inefficiency with current context parallelism (CP) designs, under variable input sequence lengths and number of input sequences per training batch.
Under the existing CP designs, CP degree is fixed throughout the entire model training, i.e., each input sequence is partitioned into an equal number of parts, which are then evenly distributed among devices.
While such partitioning guarantees memory and computation balance, communication for exchanging KV is required for every input sequence, regardless of the sequence length (Fig.~\ref{fig:moti_exp_current_cp}).

One way to improve the communication efficiency is to allow dynamic adjustment of CP and data parallelism (DP) degrees.
However, it is hard to simultaneously balance computation and memory across DP groups, since the memory requirement grows linearly with the number of assigned tokens~\cite{dao2022flashattention,korthikanti2023reducing}, while computation grows quadratically~\cite{dao2022flashattention}.
In Fig.~\ref{fig:moti_exp_dp}, the long sequence is placed on Device 0 and the shorter ones on Device 1,
achieving perfect memory balance and eliminating all CP communication, but resulting in heavy computation imbalance between devices.

Better performance can be achieved if we allow more fine-grained parallelization configurations (e.g., applying different types of parallelism for different input sequences).
In Fig.~\ref{fig:moti_exp_optimal_placement}, CP is applied to the long sequence and DP to the rest;
both memory and computation are perfectly balanced, while communication is reduced by half compared to using pure CP, leading to improved execution time in communication-bound scenarios.

\subsection{Inability to adapt to diverse attention masks}

\begin{figure}
     \centering
     \begin{subfigure}[t]{0.24\linewidth}
         \centering
         \includegraphics[width=0.9\textwidth]{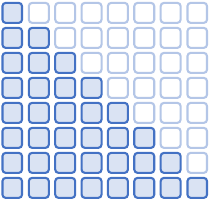}
         \caption{Causal.}
         \label{fig:attn_causal}
     \end{subfigure}
     \hfill
     \begin{subfigure}[t]{0.24\linewidth}
         \centering
         \includegraphics[width=0.9\textwidth]{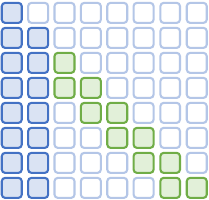}
         \caption{$\Lambda$-shaped.}
         \label{fig:attn_lambda}
     \end{subfigure}
     \hfill
     \begin{subfigure}[t]{0.24\linewidth}
        \captionsetup{justification=centering}
         \centering
         \includegraphics[width=0.9\textwidth]{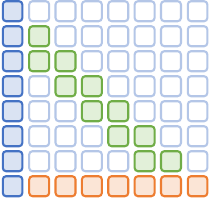}
         \caption{Causal blockwise.}
         \label{fig:attn_causal_blockwise}
     \end{subfigure}
     \hfill
      \begin{subfigure}[t]{0.24\linewidth}
        \captionsetup{justification=centering}
         \centering
         \includegraphics[width=0.9\textwidth]{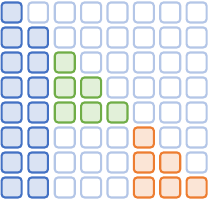}
         \caption{Shared question.}
         \label{fig:attn_shared_question}
     \end{subfigure}
     \vspace{-4mm}
        \caption{
        Attention masks used in various applications. Colors highlight different attention mask components.
        }
        \label{fig:attn_masks}
    \vspace{-4mm}
\end{figure}

\begin{figure}
     \centering
     \begin{subfigure}[t]{\linewidth}
         \centering
         \includegraphics[width=0.9\textwidth]{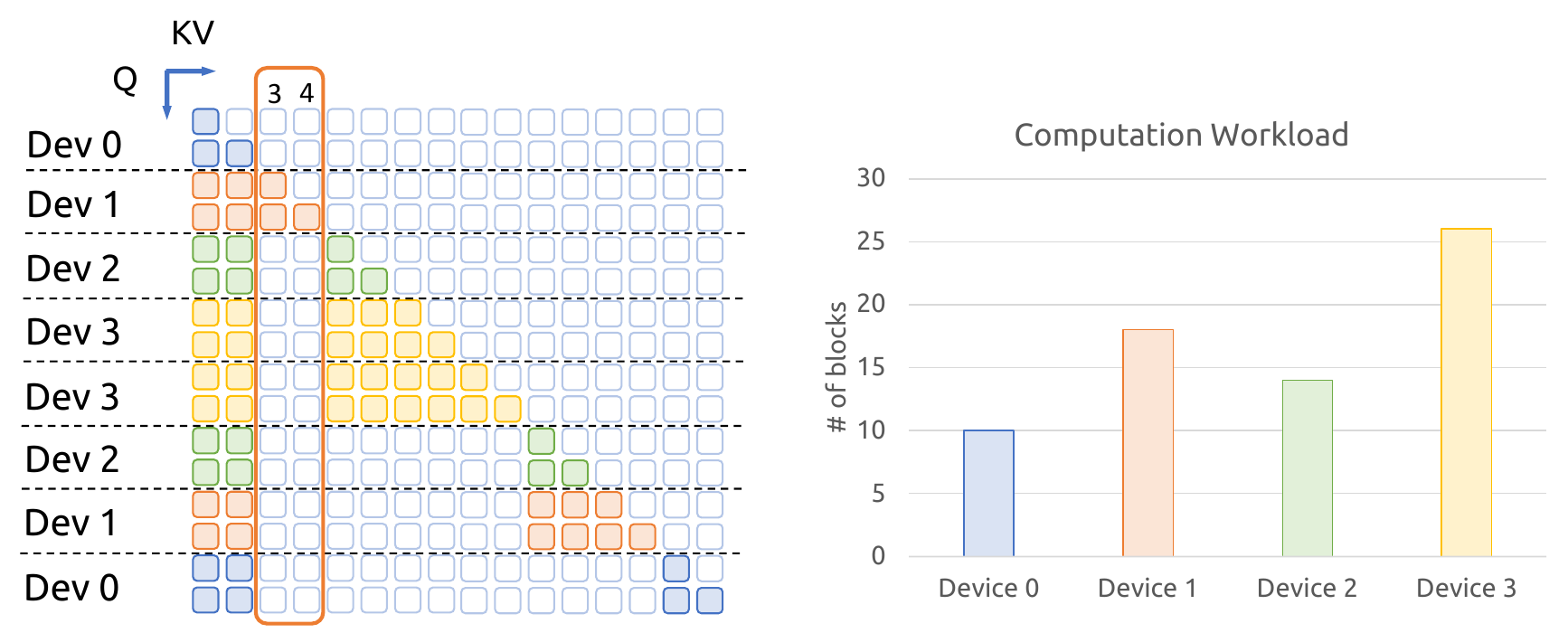}
         \caption{Colored blocks indicate the required attention between pairs of Q and KV blocks. Different colors distinguish blocks assigned to different devices.}
         \label{fig:shared_question_comp_imbal}
     \end{subfigure}
     \hfill
     \begin{subfigure}[t]{\linewidth}
         \centering
         \includegraphics[width=0.9\textwidth]{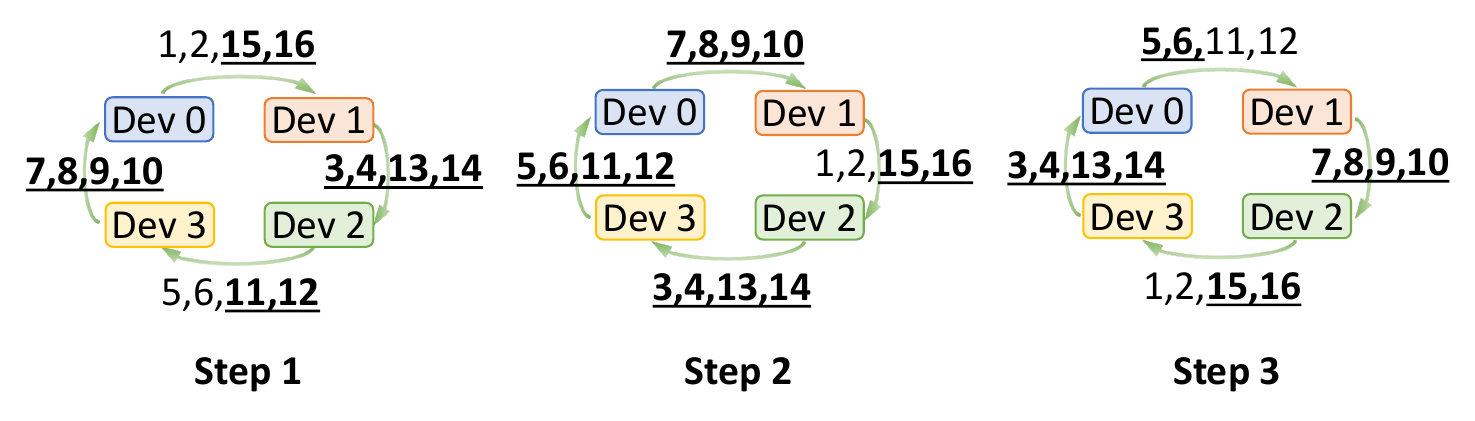}
         \caption{Communication of redundant KV blocks (those not used by the receiving device) are underlined and in bold font. Numbers on the communication arrows indicate KV block indices.}
         \label{fig:shared_question_redundant_comm}
     \end{subfigure}
    \caption{Imbalanced computation and redundant communication when applying ring attention on a shared question mask~\cite{wang2025flashmask}.}
    \label{fig:sparse_mask_inefficiency_illu}
    \vspace{-4mm}
\end{figure}

While current methods design special placement only for the causal mask (Fig.~\ref{fig:causal_placement}), more sparse attention masks are used in long-context model training and inference (Fig.~\ref{fig:attn_masks}).
Lambda($\Lambda$)-shaped mask (Fig.~\ref{fig:attn_lambda}) combines attention sink (all tokens attend to several tokens at the start of the sequence) and sliding window attention (each token only attends to a predefined number of precedent tokens), and is widely applied to reduce the total attention computation and improve training/inference speed~\cite{han2024lminfinite, xiao2024streamingllm}.
Causal blockwise mask (Fig.~\ref{fig:attn_causal_blockwise}) reduces attention computation while maintaining model performance in in-context learning (ICL) scenarios~\cite{bertsch2025incontext}:
the input is divided into blocks (each containing multiple ICL examples), attention sink and sliding window attention are applied to these blocks, and the final test example attends to all previous tokens.
In RLHF~\cite{ouyang2022RLHF} or DPO~\cite{rafailov2023dpo} training where each question may be paired with multiple candidate answers, a shared question mask (Fig.~\ref{fig:attn_shared_question}) can be used to allow answers to share the same prefix question and thus remove redundant computation~\cite{wang2025flashmask}.
Notably, in causal blockwise mask and shared question mask, the shape of the attention mask is determined not only by the model design, but also by the input data, and thus different attention masks are applied to different input batches.
These sparse masks are not supported by current context parallelism frameworks~\cite{zhu2024ringflashattn,fang2024usp,gu2024loongtrain,nvidia2024transformerengine}.

To perform distributed attention computation under these masking patterns, a simple approach is to retain current placement and communication patterns while applying attention masks to each of the local attention steps.
However, doing so may cause significant computation imbalance and communication redundancy.
In the example with a shared question mask in Fig.~\ref{fig:sparse_mask_inefficiency_illu}, current methods assign far more computation workload to device 3 (yellow blocks) than to the other devices.
Due to the mask's sparsity, the KV for token block 3-4 (columns highlighted in Fig.~\ref{fig:shared_question_comp_imbal}) is only required on Device 1, while the current communication design requires it to be transferred to all devices.
As a result, 38 out of 48 KV blocks (16 per step, over 3 steps) are redundantly communicated (see Fig.~\ref{fig:shared_question_redundant_comm}).

\subsection{Opportunities and Challenges}
To mitigate such issues caused by dynamic and flexible model input, we need fine-grained parallelism control across batches as well as for each input sequence within a batch.
Such fine-grained parallelism configuration should minimize communication while maintaining memory and computation balance.
This calls for solving the following challenges:

\noindent $\triangleright$ How to capture the effect of dynamism on context parallelism and systematically define the fine-grained parallelization configuration?

\noindent $\triangleright$ How to automatically generate optimized parallelization configurations for different batches?

\noindent $\triangleright$ How to efficiently implement such dynamic and fine-grained parallelism, with potentially different parallelization configurations for each iteration?

%% file: sections/3_system.tex
\section{Overview}\label{sec:3_sys}

We design \SystemName{}, a context parallelism framework for efficient training of large models that dynamically adapts parallelization configurations to arbitrary sequence lengths and varying attention masks. 

\subsection{Workflow}

\begin{figure}
    \centering
    \includegraphics[width=\linewidth]{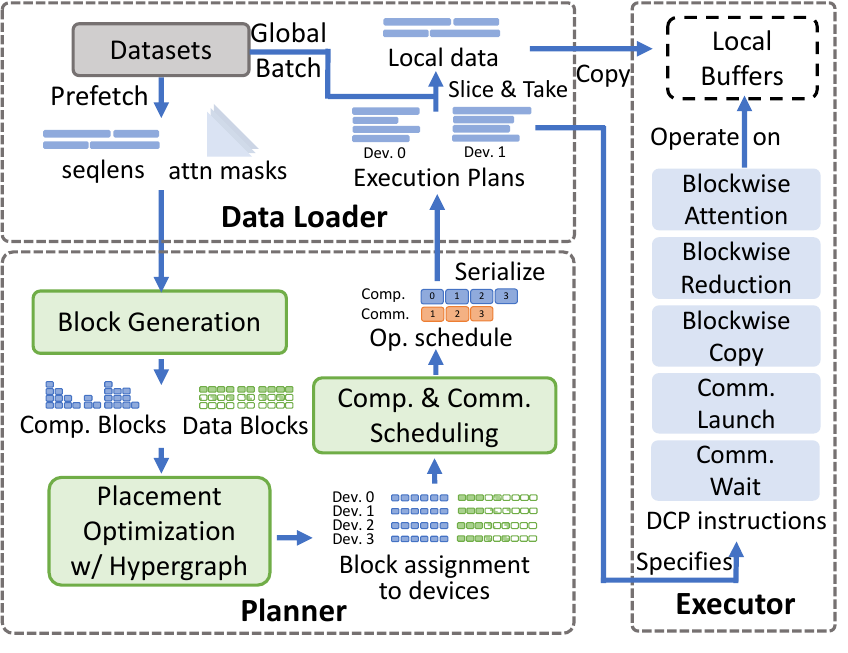}
    \vspace{-6mm}
    \caption{DCP System Overview}
    \label{fig:sec3:overview}
    \vspace{-4mm}
\end{figure}

The workflow of \SystemName{} is given in Fig.~\ref{fig:sec3:overview}. 
It consists of three key modules.

\textbf{The data-loader} pre-fetches sequence lengths and mask patterns from training datasets.
For each input batch, it performs data partitioning, generates data and computation blocks that describe context parallelism for the input batch (\S\ref{sec:block_generation}) based on sequence length and mask information, and invokes the planner.
Based on the data placement generated by the planner, it constructs the local model inputs for each device.

\textbf{The planner} solves a hypergraph partitioning problem to optimally assign data and computation to devices.
It then performs computation and communication scheduling and generates the execution plan for each device. The execution plans are serialized in the form of DCP instructions (Sec.~\ref{sec:executor}), each abstracting an elementary operation required to execute attention.
The above planning is carried out online during training and overlaps with GPU training through data pre-fetching from the datasets.

\textbf{The executor} implements the DCP instructions with high-performance kernel libraries (e.g., FlashAttention~\cite{dao2022flashattention}) or compilers like Triton~\cite{tillet2019triton}.
During each iteration, it creates a buffer storing inputs/outputs fetched from other devices and intermediate results on each device.
Then, it sequentially carries out the instructions specified by the execution plan. 

\subsection{User Interface}

\SystemName{} provides a simple user interface for easy integration by developers training large models, as given in Listing~\ref{lst:interface}. 
To use \SystemName{}, developers begin by replacing the attention implementation in their models with DCP's implementation (Lines 1-7). 
If non-causal attention masks are used, users can define a custom mask-generation function (Lines 9-12) that takes input information (available from the dataset) like sequence lengths and their composition (e.g., lengths of questions and answers when using the shared question mask).
In their training script, users construct a data-loader (Line 15) with the dataset and the mask function, and an executor (Line 18) which is shared across all model layers.
In each training iteration, users can directly get the partitioned data and corresponding execution plan from the data-loader for each device (Line 20), initialize the executor with the execution plan (Line 22), and run the model training iteration (Line 24).

\noindent
\begin{CodeListing}[DCP API]{label=lst:interface}{fontsize=\footnotesize}
# when defining models
class TransformerLayer(...):
  def forward(..., dcp_executor):
    ...
    # replace attention implementation with DCPAttn
    core_attn_out = DCPAttn.apply(dcp_executor, q, kv)
    ...

# define a mask function
def mask_fn(seqlens, ...)
  ...
  return mask

# in training script
dcp_dataloader = DCPDataloader(dataset, mask_fn)
# dcp_group is a communicator that connects all devices
# (e.g., torch.distributed.ProcessGroup)
dcp_executor = DCPExecutor(group=dcp_group)
# training iterations
for (local_data, execution_plan) in dcp_dataloader:
  # set execution plan and create buffers
  dcp_executor.prepare(execution_plan)
  # execute model
  loss = model(local_data, dcp_executor)
  ...
\end{CodeListing}

%% file: sections/4_dynamic_context_parallel.tex
\section{Dynamic Context Parallelism}\label{sec:4_dcp}

\SystemName{} advocates dynamically optimizing parallelism configurations for each training batch to best handle varying sequence lengths and attention masks, and enables switching between different configurations across training batches/iterations (i.e., \textit{dynamic context parallelism}).
To identify optimized configurations, we define a unified representation for the parallelization of a training batch.
In this approach, we divide both the data and computation of each input sequence into blocks along all parallelizable dimensions.
A parallelism configuration is then defined by assigning these blocks to devices, which in turn determines the communication pattern between devices.

\begin{figure}
     \centering
     \begin{subfigure}[c]{0.59\linewidth}
         \centering
         \includegraphics[width=\textwidth]{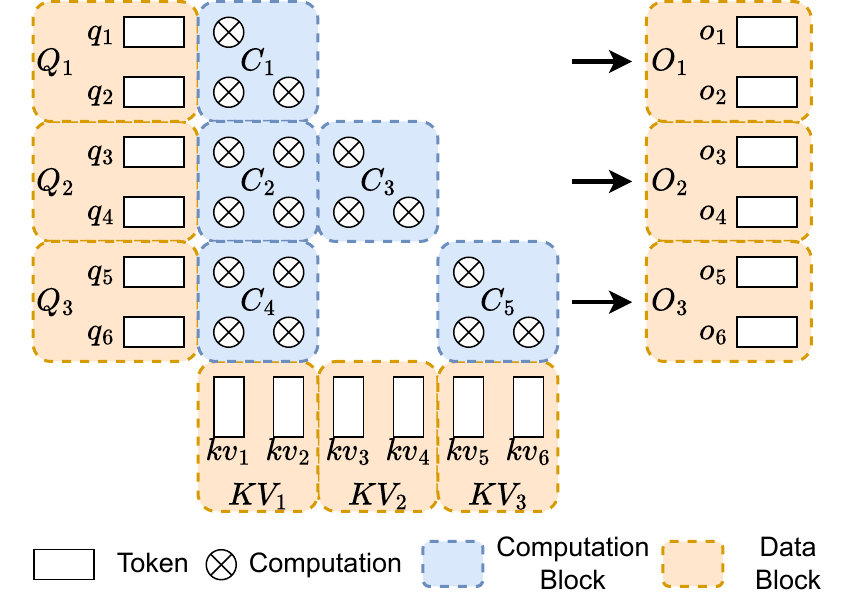}
         \caption{DCP blocks}
         \label{fig:sec4:block}
     \end{subfigure}
     \hfill
     \begin{subfigure}[c]{0.4\linewidth}
         \centering
         \includegraphics[width=\textwidth]{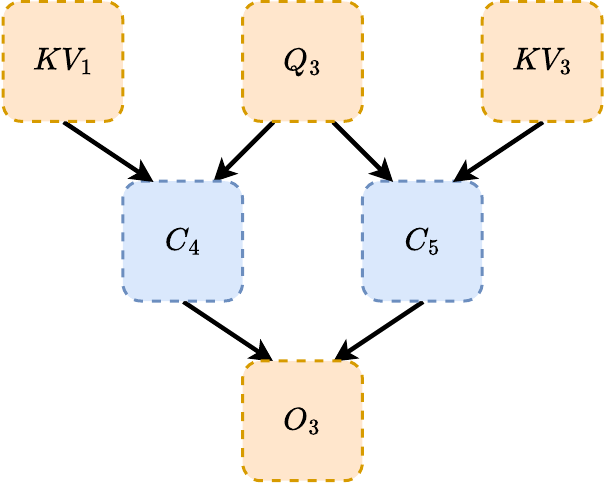}
         \caption{Block dependency.}
         \label{fig:sec4:dependency}
     \end{subfigure}
        \vspace{-4mm}
        \caption{Data and computation block partitioning.}
        \label{fig:dcp_data_and_comp_block_partitioning}
        \vspace{-4mm}
\end{figure}

\begin{figure}
     \centering
     \begin{subfigure}[t]{\linewidth}
         \centering
         \includegraphics[width=0.8\textwidth]{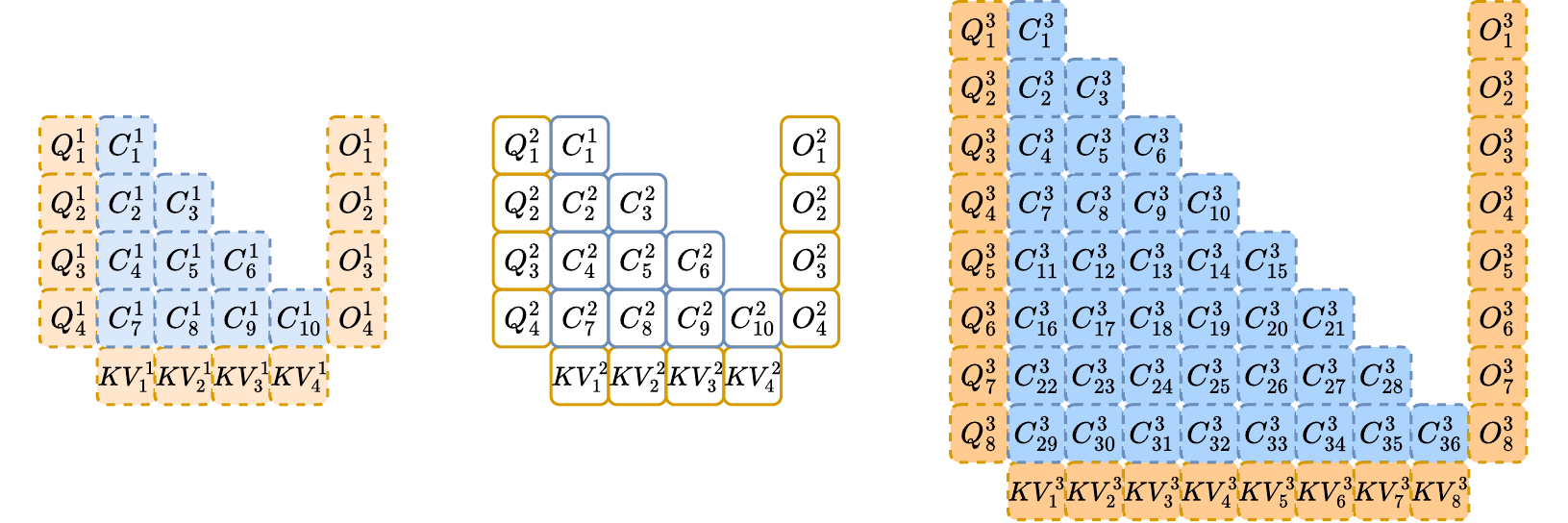}
         \caption{DCP data and computation blocks for the example in Fig.~\ref{fig:sub_optimal_ringattn} (two short and one long sequences), under causal mask. Superscript represents sequence index. Different sequences' blocks drawn with different styles. Showing blocks for a single head only.}
         \label{fig:dcp_block_causal_exp}
     \end{subfigure}
     \hfill
     \begin{subfigure}[t]{\linewidth}
         \centering
         \includegraphics[width=\textwidth]{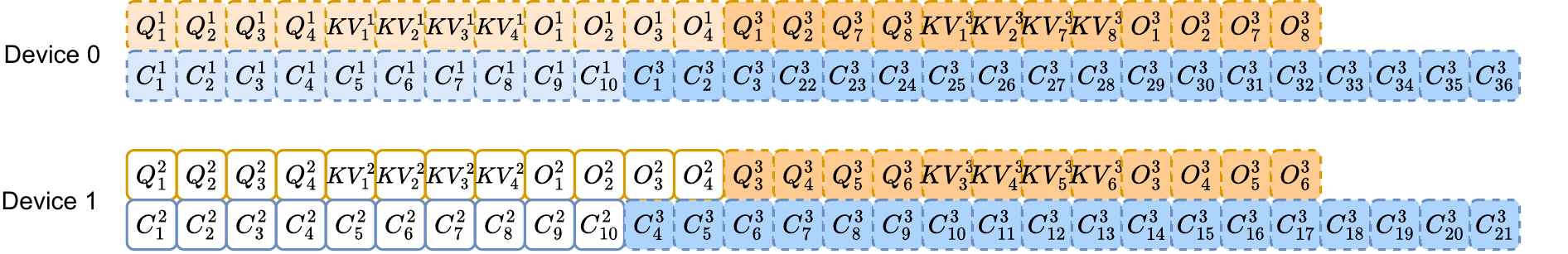}
         \caption{Block placement that corresponding to Fig.~\ref{fig:moti_exp_optimal_placement}}
         \label{fig:dcp_block_causal_exp_flexible_placement}
     \end{subfigure}
     \vspace{-4mm}
     \caption{Representing variable length sequences with DCP blocks.}
     \label{fig:dcp_blocks_causal}
\end{figure}

\begin{figure}
     \centering
     \includegraphics[width=\linewidth]{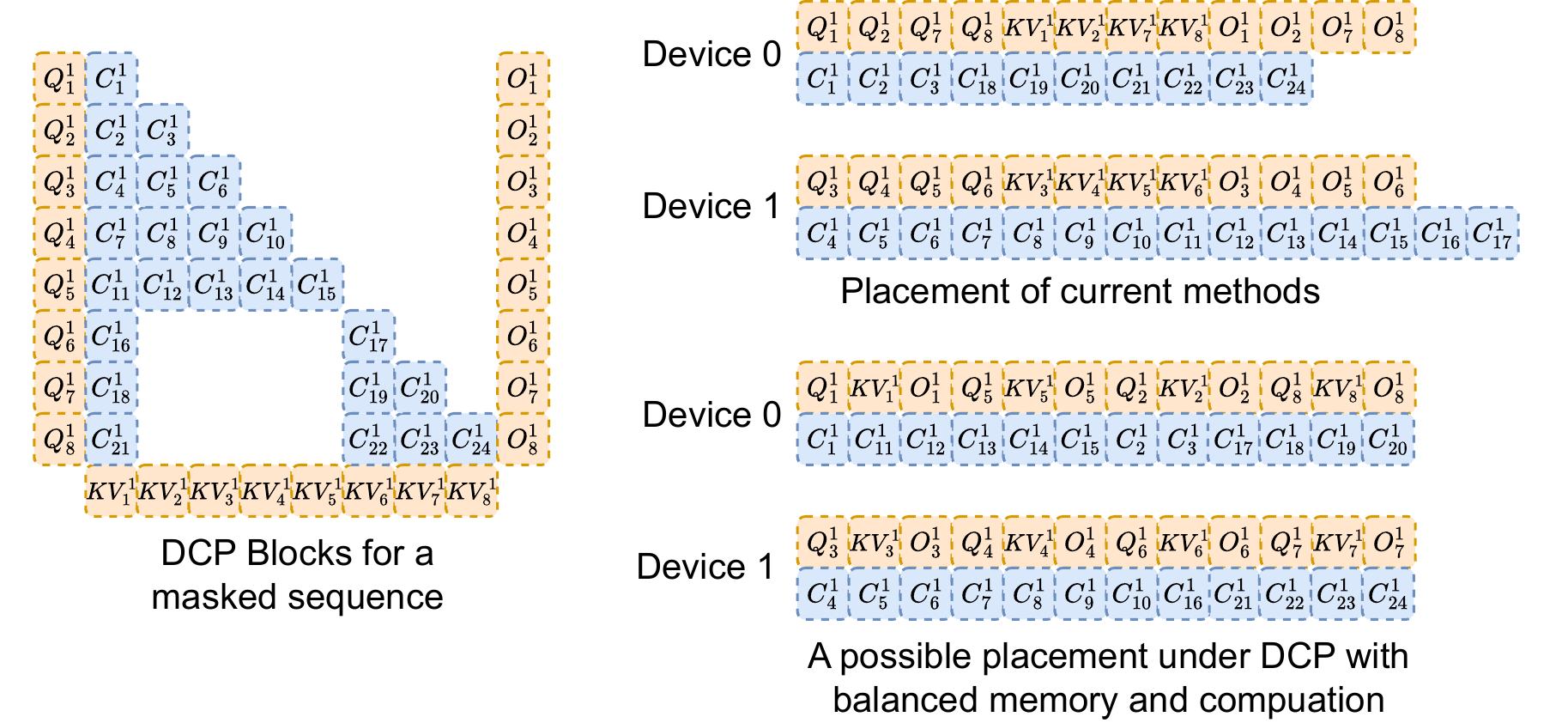}
     \vspace{-6mm}
     \caption{Example of DCP blocks for a shared question masked sequence, with two different possible placements. Showing blocks for a single head only.}
     \label{fig:dcp_block_masked_exp}
     \vspace{-4mm}
\end{figure}

\subsection{Block Generation}
\label{sec:block_generation}

For each sequence in an input batch, we partition the input and output tensors of \cd{} operators at every parallelizable dimension into contiguous slices, and denote each resulting slice as a {\em data block}.
The input tensors are $\mathbf{Q,K,V}$ and the output tensor is $\mathbf{O}$, each of shape $[H, L, D]$ (where $H$ is the number of heads, $L$ is the sequence length, and $D$ is the hidden dimension).
Each tensor is partitioned at head and sequence length dimensions, where each partitioned block is of shape $[1, \mathcal{B}, D]$, with $H \times \frac{L}{\mathcal{B}}$ blocks in total.
Here block size $\mathcal{B}$ is a hyper-parameter of the algorithm.

Computation executed by the \cd{} operator is decomposed into {\em computation blocks}, each describing the attention between a query block $\mathbf{Q_i}$ and a pair of key and value blocks $\mathbf{KV_j}$, which contributes to an output block $\mathbf{O_i}$, representing the computation of $\hat{O}_{bhij}$ in Line 5 of Listing~\ref{lst:parallel_attn_algo} (here b and h are sequence and head index for blocks $\mathbf{Q_i}$ and $\mathbf{KV_j}$).
Each of the computation blocks represents computation that can be executed in parallel.
Fig.~\ref{fig:sec4:block} shows the data and computation blocks for a single sequence of 6 tokens under a shared question mask~\cite{wang2025flashmask} (showing blocks for a single head only).
Each of $Q$, $KV$, and $O$ is partitioned into three blocks (i.e., $\mathcal{B}=2$), respectively.
For each pair of $\mathbf{Q_i}$ and $\mathbf{KV_j}$ with computation dependency (i.e., the corresponding attention mask $\mathbf{M}_{bij}$ is not all zeros), a computation block $\mathbf{C_k}$ is constructed and contributes to output block $\mathbf{O_i}$.
Fig.~\ref{fig:sec4:dependency} shows the data dependencies of computation blocks $C_4$ and $C_5$ in Fig.~\ref{fig:sec4:block}.
When multiple computation blocks contribute to the same output block, a reduction is required to aggregate the results (Lst.~\ref{lst:parallel_attn_algo}, Line 6).

We allow arbitrary device assignment of data and computation blocks, with the constraint that the blocks corresponding to $Q, KV$ and $O$ of the same tokens are placed onto the same device (since the input batch is partitioned across devices at the token level).
The placement of these $Q,KV$ and $O$ data blocks thus determines the device assignment of the corresponding tokens (i.e., we use these tokens as the input to the model replica on that device), while the placement of a computation block determines the device where the corresponding attention computation is performed.
When a computation block and the required input blocks ($Q,KV$) are assigned to different devices, communication occurs to fetch the data blocks to the computation device (and vice versa for output blocks).

With such fine-grained representation, we can easily control the parallelization configuration for each input sequence, regardless of its length.
Fig.~\ref{fig:dcp_blocks_causal} shows the data and computation blocks corresponding to the three sequence examples in Fig.~\ref{fig:sub_optimal_ringattn}.
To represent the parallelism configuration of current approaches, we assign half of data and computation blocks of each sequence to each device.
For a pure DP configuration, we assign all blocks of sequence 1, 2 to device 0, and all blocks of sequence 3 to device 1.
The configuration of Fig.~\ref{fig:moti_exp_optimal_placement} (using DP for the two short sequences while using CP for the long sequence) can be achieved by assigning all blocks of sequence 1 to device 0, all blocks of sequence 2 to device 1, and half of data and computation blocks of sequence 3 to each of the devices (Fig.~\ref{fig:dcp_block_causal_exp_flexible_placement}).

This fine-grained representation of attention computation also effectively discards unnecessary computation (i.e., where attention mask $\mathbf{M}_{bij}=0$), as those blocks will not be constructed.
The flexible block assignment makes it easy to load-balance memory and computation for masked attention.
Fig.~\ref{fig:dcp_block_masked_exp} shows the blocks generated for a sequence with shared question mask~\cite{wang2025flashmask} and possible device placements.
The blank area between computation blocks $C^1_{16}$ and $C^1_{17}$, $C^1_{18}$ and $C^1_{19}$, $C^1_{21}$ and $C^1_{22}$ are computations that are masked out.
Load balancing in DCP is achieved by assigning data/computation blocks that represent similar total data size/computation FLOPS to each device.

Such representation further facilitates modeling of communication volume incurred in any given context parallelization configuration.
Consider attention computation on an input sequence through a set of computation blocks $\mathbf{C}$. Each computation block $C_i$ consumes input data blocks $\mathbf{I_i}$ (including $Q,K,V$ blocks) and contributes to output data blocks $\mathbf{O_i}$, and each data or computation block is assigned to a device $Dev(\cdot)$.
Communication is incurred for fetching remote data blocks to the device where the computation is located, and sending output blocks to the device where the output is needed. The total communication volume under this configuration is
$\small \sum_{i=1}^{|\mathbf{C}|} \left(\sum_{j=1}^{|\mathbf{I_i}|} Size(I_{ij}) \cdot [Dev(I_{ij}) \neq  Dev(C_i)] \right.\\
\left.+\sum_{j=1}^{|\mathbf{O_i}|} Size(O_{ij}) \cdot [Dev(O_{ij}) \neq Dev(C_i)] \right)
$
where $Size(\cdot)$ denotes the size of the data block and $[x]$ is the Iverson bracket notation, equaling 1 if the condition $x$ is true, and 0, otherwise.

\begin{figure}
     \centering
     \includegraphics[width=0.7\linewidth]{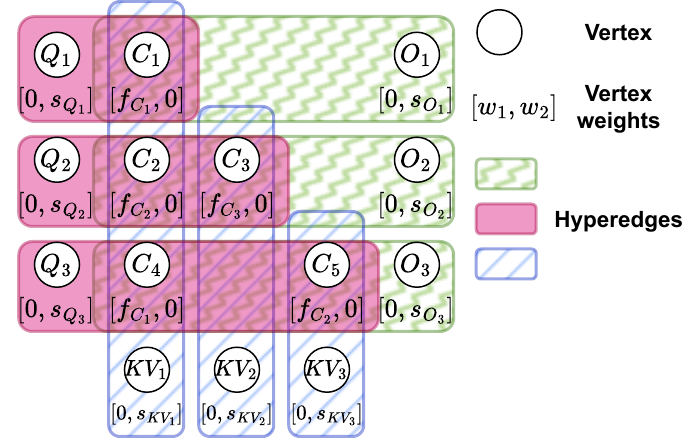}
     \caption{Hypergraph for the example in Fig.~\ref{fig:dcp_blocks_causal}.}
     \label{fig:hyper_edge}
\end{figure}

\subsection{Hypergraph partitioning for data and computation placement}

Device placement of data and computation blocks decides communication overhead in context parallelism, as well as computation load and memory consumption on devices for both \ci{} (e.g., layer normalization, MLP) and \cd{} operators.
Computation load and memory usage of \ci{} operators depend on the number of tokens processed on a device, which is in turn determined by our assignment of data blocks.
Thus we do not separately model the memory and computation of \ci{} operators; by ensuring balanced data block distribution (for attention) across devices, we automatically enable balanced \ci{} operator computation and memory consumption.

Given that intra-machine communication usually has much higher bandwidth and lower latency than inter-machine communication, it is beneficial to prioritize minimizing inter-machine communication volume.
We design an efficient \textit{hierarchical placement} scheme accordingly.
On a cluster of $X$ machines and $Y$ devices per machine, we first assign blocks to the $X$ machines, aiming to minimize cross-machine communication.
Next, we further optimize the placement of blocks on each machine onto the $Y$ devices.

For each level of placement, we describe distributed computation of an \cd{} operator on an input batch with a hypergraph $\mathbf{G}=(\mathbf{N}, \mathbf{E})$.
Hypergraphs differ from normal graphs in that each hyperedge $e\in \mathbf{E}$ can connect more than two vertices in $\mathbf{N}$, enabling the modeling of complex dependencies in applications like sparse matrix-vector multiplication~\cite{catalyurek1999hypergraphsparsemv}.
To model distributed attention, let $\mathbf{N} = \mathbf{C} \cup \mathbf{I} \cup \mathbf{O}$ denote the set of vertices, comprising computation blocks, input and output data blocks for all sequences in an input batch.
The set of hyperedges $\mathbf{E}$ is defined such that each hyperedge $e\in \mathbf{E}$ connects a data block $d$ to the set of computation blocks $\{C_i | d\in \mathbf{I_i} \text{ or } d\in \mathbf{O_i}\}$ (i.e., all computation blocks that either consume or contribute to it).
Each vertex is associated with a 2-dimensional weight, where the first weight value represents computation load and the second indicates data size.
The weight of each computation block $c\in\mathbf{C}$ (vertex $n_c$) is $\mathbf{w_{n_c}}=[f_{n_c}, 0]$ where $f_{n_c}$ represents the amount of computation, e.g., FLOPS, of the computation block.
The weight of each data block $d\in \mathbf{I_i} \text{ or }  \mathbf{O_i}$ (vertex $n_d$) is $\mathbf{w_{n_d}}=[0, s_{n_d}]$ where $s_{n_d}$ is the size of the data block.
Each hyperedge $e$ has weight $s_e$, representing data size of the data block in the hyperedge.
Fig.~\ref{fig:hyper_edge} illustrates the formulated hypergraph of Fig.~\ref{fig:sec4:block}.
Nine hyperedges are built in total, connecting each data block to its associated computation blocks.

Given $R$ devices (i.e., $R=X$ for machine-level placement and $R=Y$ for device-level placement), we divide vertices in the hypergraph into $R$ balanced partitions, $\mathbf{P_1},\ldots,\mathbf{P_r}$, by balancing the total vertex weight among partitions while minimizing a connectivity metric, and assign each partition to a device.
The connectivity metric is $\sum_{e\in \mathbf{E}} s_e(\lambda_e - 1)$ ($\lambda_{e}$ denotes the number of partitions that vertices in hyperedge $e$ span), representing the weighted sum of each hyperedge's connectivity minus one. 
When vertices connected by a hyperedge $e$ are assigned to $\lambda_e$ different devices, $s_e(\lambda_e - 1)$ represents the total communication volume of the data block in hyperedge $e$.
Therefore, the connectivity metric models the total communication volume required in the resulting hypergraph partitioning~\cite{catalyurek1999hypergraphsparsemv}.

The hypergraph partitioning and block-to-device assignment problem is:

\vspace{-5mm}
$$\min_{\mathbf{P_1,\ldots,P_r}} \sum_{e\in\mathbf{E}} s_e(\lambda_e - 1)$$
subject to 

\vspace{-7mm}
$$\mathbf{w}(\mathbf{P_i}) \preceq [1+\epsilon, 1]\odot\frac{\mathbf{w}(\mathbf{N})}{R}, \forall\ i\in [1, R]$$

\noindent where $\mathbf{w}(\mathbf{P_i})=\sum_{n\in\mathbf{P_i}}\mathbf{w_n} = [\sum_{n\in\mathbf{P_i}}f_{n_c}, \sum_{n\in\mathbf{P_i}}s_{n_d}]$ is the total vertex (computation and data) weight in $\mathbf{P_i}$, $\preceq$ denotes less or equal in both weight dimensions, and $\odot$ denotes element-wise multiplication.
$\epsilon$ is a small positive value specifying the computation imbalance tolerance (we always try to make data blocks as balanced as possible).
The constraint ensures computation and memory balance across partitions.
This balanced hypergraph partitioning problem is NP-hard~\cite{garey1976hypergraphnphard}, with efficient heuristics available (e.g., multi-level partitioning~\cite{catalyurek1996multilevel}), implemented by off-the-shelf solvers such as PaToH~\cite{ccatalyurek2011patoh} and KaHyPar~\cite{schlag20kahypar}.

\subsection{Computation \& communication scheduling}
\label{sec:comp_comm_scheduling}

Placement of data and computation blocks determines the necessary communication between devices.
However, sequentially executing assigned computation blocks on each device, along with their associated communication, may not saturate hardware resources, including both computing power and bandwidth.
To enhance hardware utilization and minimize communication overhead, we advocate a multi-division execution schedule, by grouping computation block on each device into divisions and overlapping the computation of one division with the communication of the next, where communication and computation for each division is executed with fused kernels.

Computation blocks from a training batch can be computed in any order, as they are parallelizable computation units divided along parallelizable dimensions of the \cd{} operator.
However, block division determines communication scheduling, as each computation block may be associated with different amounts of communication from distinct devices.
Ideally, we want divisions on every device to have balanced computation and communication load, as well as balancing among devices.
Finding such divisions is equivalent to solving a multi-dimensional assignment problem (i.e., assigning computation blocks into $T$ divisions while minimizing the maximum computation/communication on each device), which is NP-complete~\cite{frieze1983complexity}.
We devise a greedy heuristic to find balanced divisions (Listing~\ref{lst:block_scheduling}).

For each device, we calculate the total required communication volume of its assigned computation blocks (total size of data sent to/received from other devices), and divide the workload such that each division accounts for $\frac{1}{T}$ of this total communication (Lines 12-14 in Listing~\ref{lst:block_scheduling}).
We schedule all computation blocks that do not require communication (Line 16-20) into the first division on each device.
Then the device with the least computation load is selected and we schedule its second division by going through all the rest computation blocks on this device: if scheduling a block into this division causes the communication to exceed the per-stage requirement (i.e., $\frac{1}{T}$ of total communication), we defer the block to the following division; otherwise, we schedule the block into this division.
We repeat the above procedure to schedule the second division on all devices.
Similarly, the third, fourth, $\ldots$, $T-1$th divisions are scheduled consecutively on each device (Lines 28-35).
We schedule all remaining blocks into the final $D$th division, regardless of their communication volumes (Lines 21-26).
If the output block of a computation block on a device is placed on another device, we perform output transfer after all divisions and corresponding output reduction are performed on this device (Lines 36-38).

The communication and computation of each division, as well as the associated reduction operations, are serialized into an execution plan that contains a list of DCP instructions (each representing a fused operation) and is ready to be consumed by the executor. 

\noindent
\begin{CodeListing}[Block Scheduling Pseudo-code]{label=lst:block_scheduling,float=t}{fontsize=\scriptsize,tabsize=2}
# Inputs:
# hg: partitioned hypergraph containing computation and data
#     blocks, with each block assigned to a device
# T: number of divisions
# devices: all devices participating in DCP
# Output: divs[division][device][comp or comm]
#   comp blocks and communication in each division 
#   on each device
def schedule(hg, T, devices):
  # comm_requirements[d1][d2]:
  #   the total amount of data transfer from d2 to d1 
  comm_requirements = calc_required_comms(hg, devices)
  for (d1, d2), comm_size in comm_requirements:
    per_div_comm_limit[d1][d2] = comm_size / T
  for i in {0,...,T-1}:
    if i == 0:
      # first division
      for d in devices:
        divs[i][d][comp] = comp_with_no_comm(hg, d)
        divs[i][d][comm] = None
    elif i == T-1:
      # last division
      for d in devices:
        remaining_blocks = get_remaining_blocks(hg, divs, d)
        divs[i][d][comp] = remaining_blocks
        divs[i][d][comm] = get_comms(hg, remaining_blocks)
    else:
      while exists schedulable blocks:
        d = get_device_with_min_comp_load(hg, divs[i])
        for comp_block in get_remaining_blocks(hg, divs, d):
          if scheduling comp_block violates per_div_comm_limit:
            continue
          else:
            divs[i][d][comp].append(comp_block)
            divs[i][d][comm].append(get_comms(comp_block))
  # schedule communication for output blocks, if required
  for d in devices:
    divs[N][d][comm] = get_output_comms(hg, d)
  return divs
\end{CodeListing}

%% file: sections/5_executor.tex
\section{Executor}
\label{sec:executor}

DCP's executor on each device adopts a block-centric design, consisting of two key components: block buffers and the instructions that operate on them.

\textbf{Block buffers} reside in GPU memory and hold all data blocks used by this device, including local input ($Q$, $KV$) and local output ($O$), data fetched from other devices, and intermediate results pending for reduction.
To reduce memory fragmentation, a single contiguous buffer is used for all data blocks of the same type (e.g., all $Q$ blocks).
Each data block is thus identified by its type and its index into the buffer.
The buffer index of each data block is determined during computation/communication scheduling (\S\ref{sec:comp_comm_scheduling}) by a buffer manager that keeps track of which buffer indices are occupied.
We maximally reuse buffer indices that contain no longer needed blocks to minimize the total buffer size.

We abstract 5 types of DCP instructions for elementary operations in distributed attention execution:

\noindent$\bullet$ \textbf{Blockwise Attention} executes fused and masked attention 
(Listing~\ref{lst:parallel_attn_algo} Line 5) specified by all computation blocks in a division.
Especially, it takes a list of query, key-value, and output tuples as input and performs attention on corresponding blocks.
We base our attention implementation on FlashAttention~\cite{dao2022flashattention}.
Since the input and output blocks may not be contiguous in memory, we modify FlashAttention and enable all input/output blocks to be specified by the block buffer's starting address and offsets of blocks recorded in block tables (similar to PagedAttention~\cite{kwon2023pagedattention}).
We support various attention masks via arrays specifying the index ranges each token should attend to, with the limitation of at most two ranges for each token (for simplicity of implementation).
More flexible implementation is possible by adopting methods for optimizing sparse attention kernels like FlexAttention~\cite{dong2024flexattn} and FlashMask~\cite{wang2025flashmask}, which is orthogonal to this paper.

\noindent$\bullet$ \textbf{Blockwise Reduction} takes multiple attention output blocks as input and performs fused update and reduction operation (Listing~\ref{lst:parallel_attn_algo}, Line 6) on these blocks.
We implement this kernel with Triton~\cite{tillet2019triton}.

\noindent$\bullet$ \textbf{Blockwise Copy} takes multiple input or output data blocks as input and performs fused GPU memory copy on a single device.
This operation is used for managing buffers during DCP execution and is implemented with Triton~\cite{tillet2019triton}.

\noindent$\bullet$ \textbf{Comm.~Launch} takes a list of data blocks and asynchronously launches the communication to transfer these data blocks between devices.
It is implemented with PyTorch's P2P communication primitives~\cite{ansel2024pytorch2}, which internally use NCCL~\cite{nvidia2024nccl} as backend.

\noindent$\bullet$ \textbf{Comm.~Wait} instructs the GPU to wait for a previously launched communication.

An execution plan consists of a sequence of such instructions with corresponding arguments. At runtime, the instructions in an execution plan are sequentially executed by the executor.

%% file: sections/6_other_impl_details.tex
\section{Other Implementation Details}
\SystemName{}'s core modules (dataloader, planner, executor) are implemented with 14k LOCs in Python, with an additional 300 LOCs for accelerating the computation/communication planning algorithm in C++.

\subsection{Overlapping planning with model execution}
The DCP dataloader pre-fetches input information for the planner to produce an execution plan for each training iteration asynchronously, well before the actual training iteration starts.
Specifically, developers can define $\kappa$, the number of look-ahead iterations.
The dataloader tries to ensure that when executing iteration $i$, the planning for iterations $i$ through $i+\kappa$ is finished.
Whenever this condition is not met, it pre-fetches input information for a new iteration and invokes a planner instance for planning.
The planning of iterations $i$ through $i+\kappa$ can be conducted in parallel.
For multi-machine distributed training, we assign the execution planning (i.e., taking sequence lengths and attention masks as input and generating execution plans for each device) of different iterations to different machines.
On each machine, the planning for different training iterations is parallelized to run on different CPU cores.
The resulting execution plans are distributed to each device via a distributed key-value store (e.g., Redis~\cite{redis2025redis}) which is located in host memory in one of the machines.
This parallel execution planning using CPU resources greatly reduces the planning overhead, allowing us to overlap planning with actual model execution effectively.

\subsection{Combining with other parallelisms}

Data parallel training of different input sequences is included in parallelization configurations that can be produced by DCP.
Tensor parallelism is orthogonal to DCP's parallelization configurations, but it too partitions the head dimension of all attention input and output tensors (Q, KV, O).
When jointly used with tensor parallelism, DCP's head dimension size should be divided by the tensor parallel degree (i.e., the number of devices that partition a tensor), and the same execution plan is shared among different tensor parallel groups.
Pipeline parallelism is also orthogonal to DCP.
It splits different model layers across stages, but each stage can still use context parallelism, where DCP’s optimizations can be applied.
When used in conjunction with TP and PP, DCP should occupy the device ranks traditionally assigned to DP and CP (e.g., following the TP-CP-DP-PP rank assignment order in Megatron-LM~\cite{shoeybi2020megatronlm}).
For example, TP can be applied among consecutive ranks within a node to mitigate its high communication cost, followed by DCP which incurs less communication overhead than TP but more than PP, and finally, PP can be applied among distant ranks.

%% file: sections/7_evaluation.tex
\section{Evaluation}
We conduct extensive experimental evaluation of DCP in two steps: micro-benchmarking of the attention operator, and end-to-end evaluation of whole model training performance.
We also evaluate the detailed performance of DCP across different parameters.

\subsection{Attention micro-benchmarking}
\label{sec:microbenchmark}

\begin{figure*}
     \centering
     \begin{subfigure}[t]{0.49\linewidth}
         \centering
         \includegraphics[width=\textwidth]{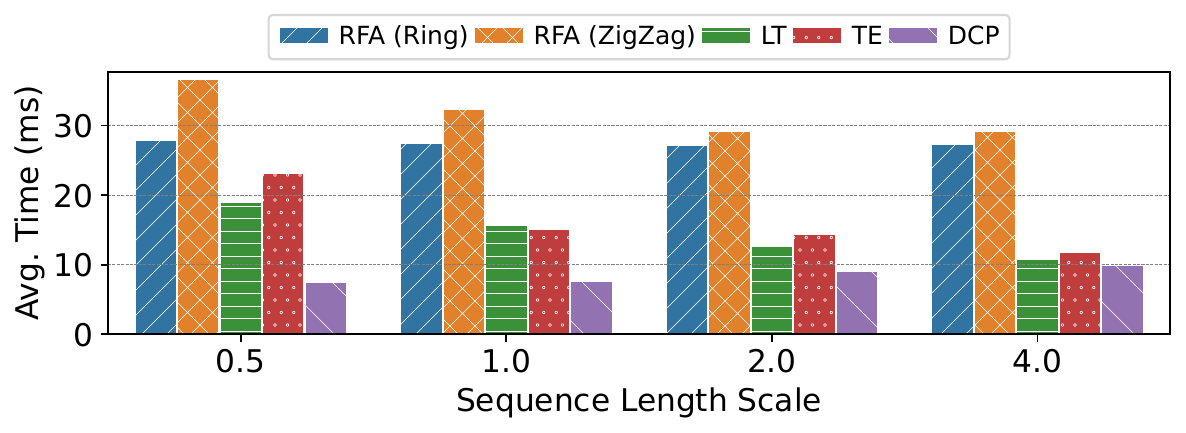}
         \vspace{-6mm}
         \caption{Attention Forward (FW)}
         \label{fig:microbenchmark_causal_FW}
     \end{subfigure}
     \hfill
     \begin{subfigure}[t]{0.49\linewidth}
         \centering
         \includegraphics[width=\textwidth]{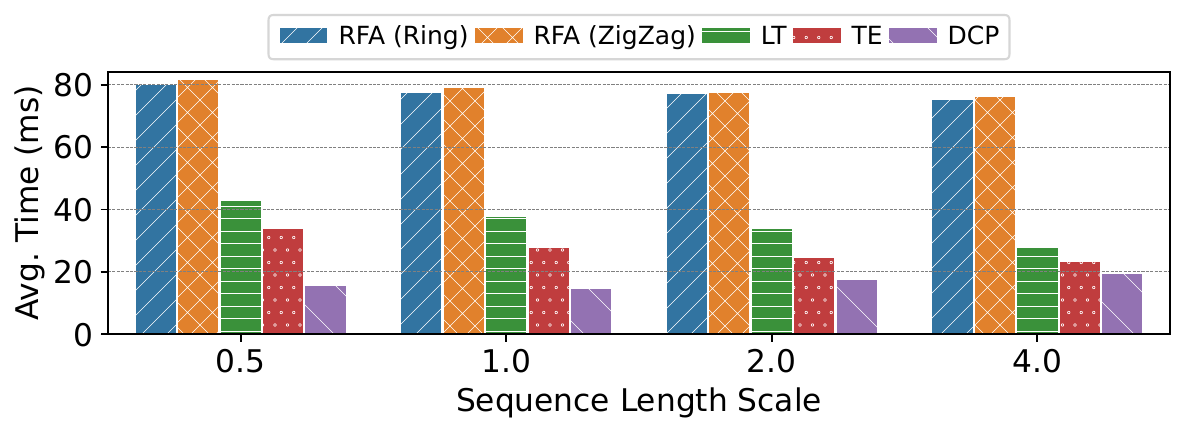}
         \vspace{-6mm}
         \caption{Attention Backward (BW)}
         \label{fig:microbenchmark_causal_BW}
     \end{subfigure}
     \vspace{-4mm}
    \caption{Micro-benchmark attention performance with causal mask.}
    \label{fig:microbenchmark_causal}
    \vspace{-4mm}
\end{figure*}

\begin{figure*}
     \centering
     \includegraphics[width=\textwidth]{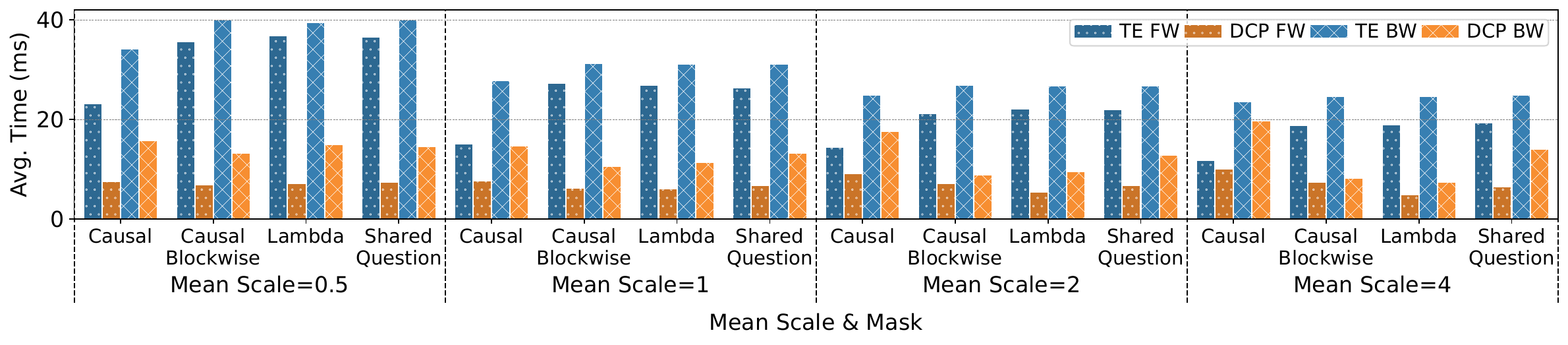}
     \vspace{-8mm}
    \caption{Micro-benchmark attention performance under different attention masks.}
    \label{fig:microbenchmark_masks}
\end{figure*}

We first evaluate the distributed attention performance of DCP in detail.

\vspace{1mm}
\noindent \textbf{Testbed.}
We conduct the micro-benchmarking experiments on 4 Amazon EC2 \texttt{p4de.24xlarge} instances.
Each \texttt{p4de} instance is equipped with 8 NVIDIA A100 (80GB) GPUs and 96 vCPU cores.
The 8 GPUs in each instance are connected by NVSwitch with 600GB/s bidirectional bandwidth.
Inter-instance communication is supported by 4x100 Gbps NICs on each instance with EFA~\cite{aws2022efa} enabled.

\vspace{1mm}
\noindent \textbf{Baselines.}
We compare \SystemName{} with state-of-the-art long-context model training frameworks: (i) RingFlashAttention (RFA)~\cite{zhu2024ringflashattn}, which parallelizes attention computation only at the sequence length dimension and supports two different input placements: Ring, which splits each input sequence into $R$ blocks ($R$ is the number of devices) and places the $r$th block of the sequences onto the $r$th device;
ZigZag, which splits each sequence into $2R$ blocks and uses a zig-zag pattern for block-to-device assignment (such a placement pattern is also used by the following two baselines (\S\ref{sec:background_cp}).
(ii) LoongTrain (LT)~\cite{gu2024loongtrain} parallelizes attention at both head and sequence length dimensions and does not support variable-length input sequences, so we pad the sequences to the longest sequence length in each batch.
It requires specifying the size of the inner ring for its double ring attention (a communication schedule that aims to improve NIC utilization), and we experiment under inner-ring sizes 1,2,4,8 and report the best result.
(iii) TransformerEngine (TE)~\cite{nvidia2024transformerengine} also parallelizes attention at both head and sequence length dimensions and lacks variable-length input support when parallelizing at both dimensions, but can be easily extended to accommodate it.
None of the baselines support attention masks other than causal masks.
To facilitate a direct comparison, we add this support to TransformerEngine by (pre-)computing a local mask for each of its computation steps and use DCP's masked attention kernels, without changing its communication pattern.

\vspace{1mm}
\noindent \textbf{Hyper-parameters.}
DCP requires a few hyper-parameters: block size $\mathcal{B}$ (that we use to partition the sequence length dimension), the number of divisions for each input batch in computation/communication planning (\S\ref{sec:comp_comm_scheduling}), 
and the computation imbalance tolerance ($\epsilon$).
We empirically fix the number of divisions to 4 since it yields overall good performance in our test scenarios.
We search through block sizes 512, 1024, 2048, 4096 and report the best performance, and use inter-node $\epsilon = 0.4$ and intra-node $\epsilon = 0.1$ in all our experiments (except in the ablation study in Sec.~\ref{sec:ablation}).
We use KaHyPar~\cite{schlag20kahypar} as our hypergraph partitioning solver.

\vspace{1mm}
\noindent \textbf{Dataset.}
We use the LongDataCollections~\cite{togethercomputer2024longdatacollections} dataset, a compilation of common long-context datasets designed for long input understanding tasks.
It exhibits skewed, long-tailed sequence length distributions similar to those found in larger pre-training datasets like the Pile~\cite{gao2020thepile} (Fig.~\ref{fig:dataset_seqlen_distribution}).
Such a sequence length distribution pattern is also observed in other studies~\cite{wang2025wlbllm,ge2025bytescale}.
Additionally, to better understand the influence of the number of short sequences on the performance, we obtain four variations by multiplying the length of each sequence by 0.5, 1 (no scaling), 2 and 4.
We use a global batch size of 131072 tokens, and set the maximally allowed sequence length to the same value.
We report the average attention execution time over the first 200 batches.

\vspace{1mm}
\noindent \textbf{Attention Masks.}
\label{para:attn_mask_spec}
We implement four types of attention masks: causal mask, causal blockwise mask~\cite{bertsch2025incontext}, lambda ($\Lambda$) mask~\cite{han2024lminfinite,lin2024infinitellm}, and the shared question mask~\cite{wang2025flashmask,ouyang2022RLHF}.
Lambda mask uses 64 attention sink tokens and a window size of 4096.
Causal blockwise masking uses a fixed block size of 256, a sliding window of 2 blocks, and a single block for both the attention sink and test sample.
Shared question mask assumes a single shared question with 4 different answers, each taking
up 20\% of the entire sequence length.
The same mask specification is used in the end-to-end evaluation as well.

\vspace{1mm}
\noindent \textbf{Attention Op.~Spec.} We use GQA~\cite{ainslie2023gqa}, with 8 total heads for Q, 2 groups for KV and head dimension 128 (corresponding to a 32 head, 8 KV group attention operation using 4-way tensor parallelism on the head dimension).
All 32 GPUs are used in context parallelism.
For baselines that support parallelizing both head and sequence length dimensions, we set the head parallelization size to 2 (the number of KV groups) to minimize their communication.

Fig.~\ref{fig:microbenchmark_causal} shows the average execution time of the forward pass and the backward pass of the attention operator when using the causal attention mask, under different sequence length scalings.
DCP achieves the highest speed-up in most cases.
DCP dynamically computes the parallelization configurations according to input sequence lengths and can place an entire short sequence on one device to minimize communication, while the baselines require nearly the same communication for all batches; thus DCP's speed-up is highest when there are more short sequences (at sequence length scale 0.5), 2.45x (considering both forward and backward) compared to the next best baseline (LT).
As the sequence length scale increases, DCP's attention time becomes close to LT or TE's, since there are little opportunities for DCP's parallelization optimization as compared to baselines, for batches consisting of only a small number of long sequences.
RFA performs the worst overall, as it does not support parallelizing along the head dimension, resulting in significantly higher communication costs.
LT's performance improves with increasing sequence length scale due to the larger amount of padding in batches of variable-length sequences.
Similarly, TE's performance improves as sequence length scale increases, although it does not rely on padding.
We suspect this is due to the setting being communication bound (attention computation time is small), while overheads (e.g., reordering tensors between head and ring parallelization, constructing attention arguments) decrease with the number of sequences.

Fig.~\ref{fig:microbenchmark_masks} shows the performance of DCP and TE under different attention masks.
On sparse masks, DCP significantly outperforms TE at up to 3.77x, since DCP removes redundant communication with its fine-grained block partitioning and placement.
The speed-up is more significant on causal blockwise and lambda masks compared with shared question mask, since the former two exhibit more sparsity.

\subsection{End-to-end evaluation}
\label{sec:e2e_eval}

\begin{figure*}
% \captionsetup[subfigure]{justification=centering}
     \centering
     \begin{subfigure}[t]{0.24\linewidth}
         \centering
         \includegraphics[width=\textwidth]{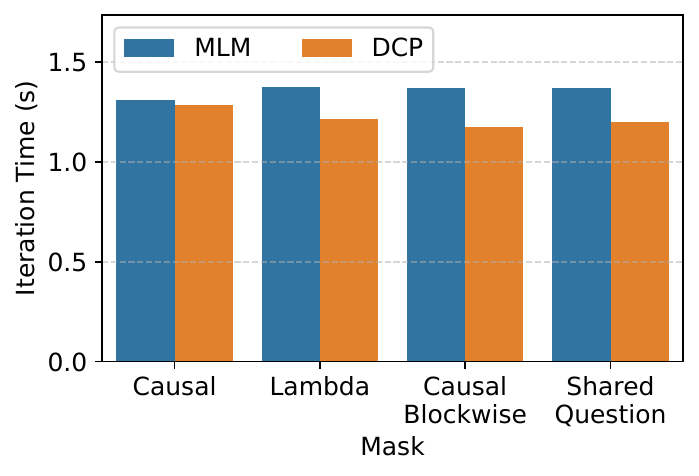}
         \vspace{-6mm}
         \caption{Max Seq. Len. 16384.}
         \label{fig:e2e_longalign_16384}
     \end{subfigure}
     \hfill
     \begin{subfigure}[t]{0.24\linewidth}
         \centering
         \includegraphics[width=\textwidth]{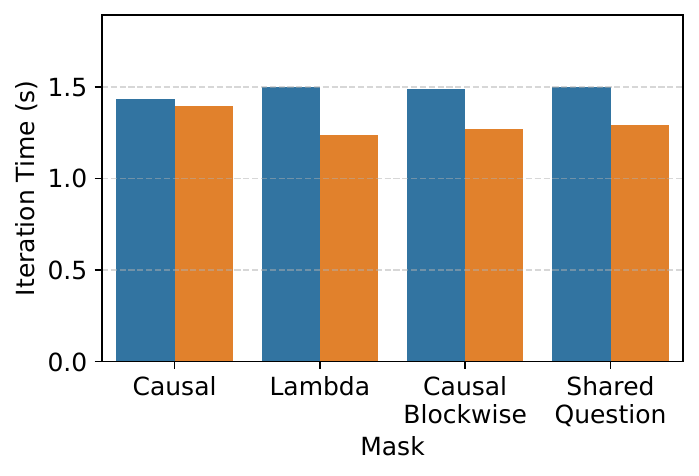}
         \vspace{-6mm}
         \caption{Max Seq. Len. 32768.}
         \label{fig:e2e_longalign_32768}
     \end{subfigure}
     \hfill
     \begin{subfigure}[t]{0.24\linewidth}
         \centering
         \includegraphics[width=\textwidth]{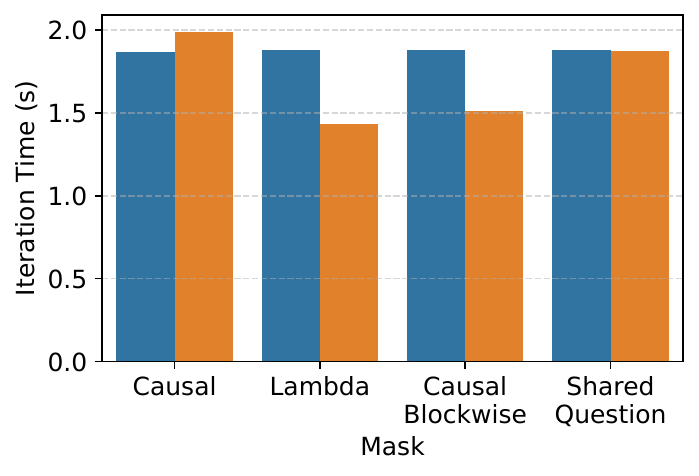}
         \vspace{-6mm}
         \caption{Max Seq. Len. 65536.}
         \label{fig:e2e_longalign_65536}
     \end{subfigure}
     \hfill
     \begin{subfigure}[t]{0.24\linewidth}
         \centering
         \includegraphics[width=\textwidth]{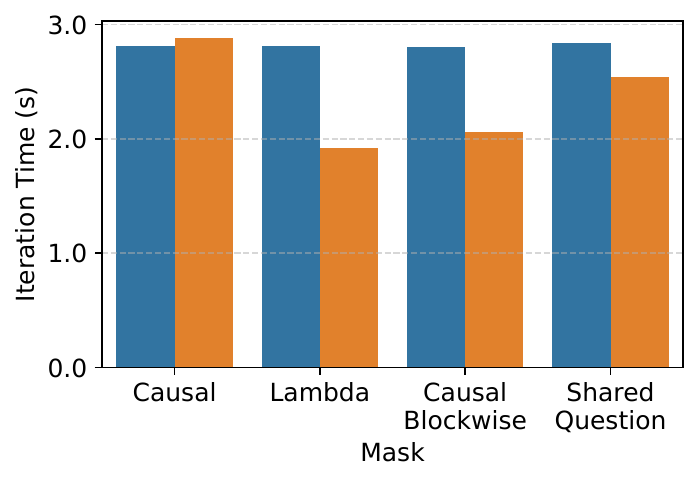}
         \vspace{-6mm}
         \caption{Max Seq. Len. 131072.}
         \label{fig:e2e_longalign_131072}
     \end{subfigure}
    \vspace{-2mm}
    \caption{End-to-end training performance on the LongAlign dataset.}
    \vspace{-2mm}
    \label{fig:e2e_longalign}
\end{figure*}

\begin{figure*}
% \captionsetup[subfigure]{justification=centering}
     \centering
     \begin{subfigure}[t]{0.24\linewidth}
         \centering
         \includegraphics[width=\textwidth]{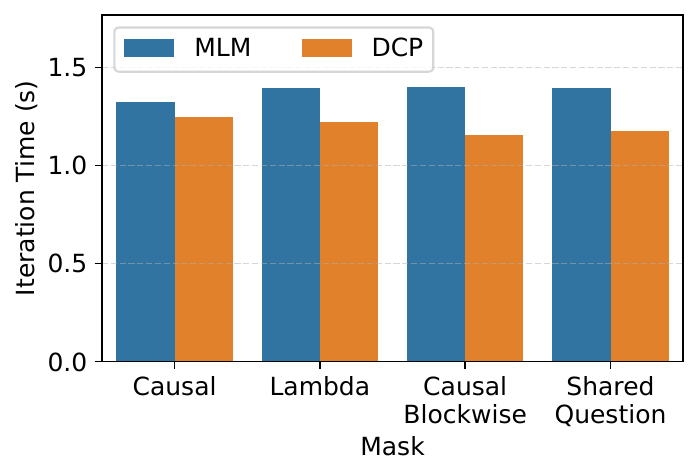}
         \vspace{-6mm}
         \caption{Max Seq. Len. 16384.}
         \label{fig:e2e_ldc_16384}
     \end{subfigure}
     \hfill
     \begin{subfigure}[t]{0.24\linewidth}
         \centering
         \includegraphics[width=\textwidth]{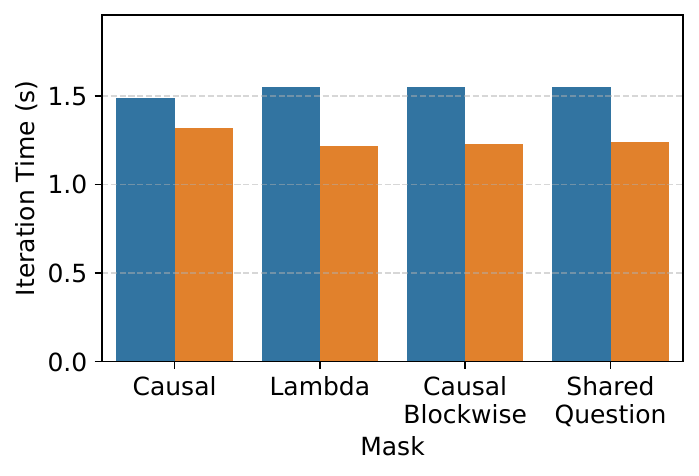}
         \vspace{-6mm}
         \caption{Max Seq. Len. 32768.}
         \label{fig:e2e_ldc_32768}
     \end{subfigure}
     \hfill
     \begin{subfigure}[t]{0.24\linewidth}
         \centering
         \includegraphics[width=\textwidth]{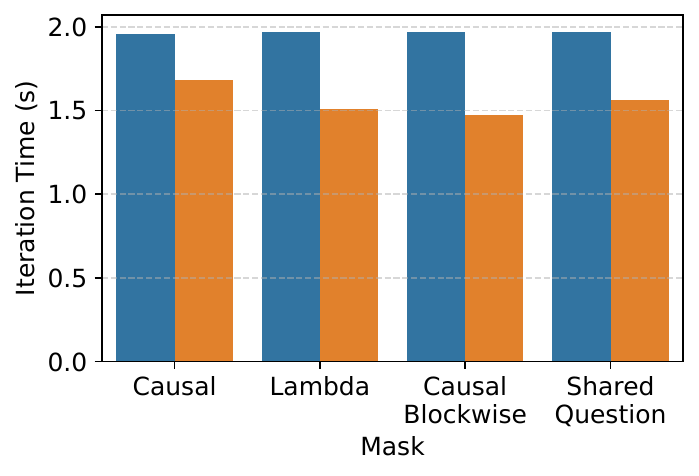}
         \vspace{-6mm}
         \caption{Max Seq. Len. 65536.}
         \label{fig:e2e_ldc_65536}
     \end{subfigure}
     \hfill
     \begin{subfigure}[t]{0.24\linewidth}
         \centering
         \includegraphics[width=\textwidth]{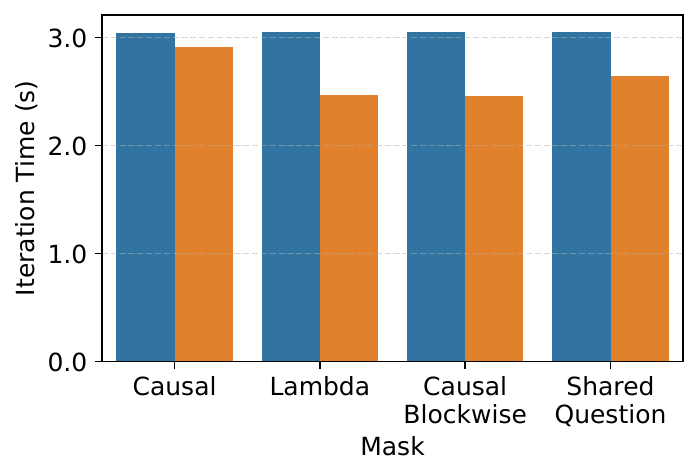}
         \vspace{-6mm}
         \caption{Max Seq. Len. 131072.}
         \label{fig:e2e_ldc_131072}
     \end{subfigure}
     \vspace{-2mm}
    \caption{End-to-end training performance on the LongDataCollections dataset.}
    \vspace{-2mm}
    \label{fig:e2e_ldc}
\end{figure*}

We first evaluate end-to-end model training performance with DCP.

\vspace{1mm}
\noindent \textbf{Testbed.}
We conduct end-to-end experiments on eight Amazon EC2 p4de.24xlarge instances (64 GPUs in total).
While the instances are the same as those used in the micro-benchmarks, a larger scale is employed to accommodate the high memory demands of end-to-end model training.

\vspace{1mm}
\noindent \textbf{Baseline.}
Our models are implemented in Megatron-LM (MLM)~\cite{shoeybi2020megatronlm}, a highly regarded framework for training models with state-of-the-art performance.
We replace the attention module in Megatron-LM with DCP's executor.
For the baseline, Megatron-LM natively supports context parallelism through TransformerEngine~\cite{nvidia2024transformerengine}; we use our enhanced TE (supporting variable-length inputs and attention masks) in its place as the baseline.

\vspace{1mm}
\noindent \textbf{Dataset.}
In addition to the LongDataCollections~\cite{togethercomputer2024longdatacollections} dataset used in \S\ref{sec:microbenchmark}, we experiment on the LongAlign~\cite{bai2024longalign} dataset, which is used for long-context LLM alignment (post-training).
It has longer average sequence lengths and fewer short sequences compared to LongDataCollections, while exhibiting similar sequence length distribution patterns (Fig.~\ref{fig:dataset_seqlen_distribution}).

\vspace{1mm}
\noindent \textbf{Model Spec.}
We implement a GPT (8B)~\cite{radford2019gpt2} model with 32 layers, hidden size 4096, 32 heads, 8 KV groups, head dimension 128 and FFN hidden size 14336 (corresponding to the setup of Llama3-8B~\cite{dubey2024llama3herdmodels}).
We use 4-way tensor parallelism within each instance, and 16-way context parallelism among the rest of the devices. 

\vspace{1mm}
Fig.~\ref{fig:e2e_longalign} and Fig.~\ref{fig:e2e_ldc} show the per-iteration training time on LongAlign and LongDataCollections datasets, respectively.
Overall, DCP achieves up to 1.16x speed-up when using the causal mask, and 1.46x under sparse masks.
The speed-up appears smaller compared to those in micro-benchmark experiments since the execution time of \ci{} operators and the time needed for gradient synchronization is similar for both DCP and MLM baseline.
For the causal mask, the speed-up is higher on the LongDataCollections dataset, since it has more short sequences.
On the LongAlign dataset, DCP can underperform MLM under the causal mask when the maximum sequence length is large.
We attribute this to less overlap between computation and communication (analyzed in \S~\ref{sec:decomposition}).
The speed-up under causal mask is also higher at smaller maximum sequence lengths, since shorter maximum sequence lengths exclude long sequences and lead to higher percentages of short sequences in batches.
We also observe consistent speed-ups with sparse attention masks, with higher speed-ups under lambda and causal blockwise masks, aligning with observations in \S\ref{sec:microbenchmark}.

\subsection{Further Performance Analysis}
\label{sec:ablation}

\begin{figure}
     \centering
     \begin{subfigure}[t]{0.49\linewidth}
         \centering
         \includegraphics[width=\textwidth]{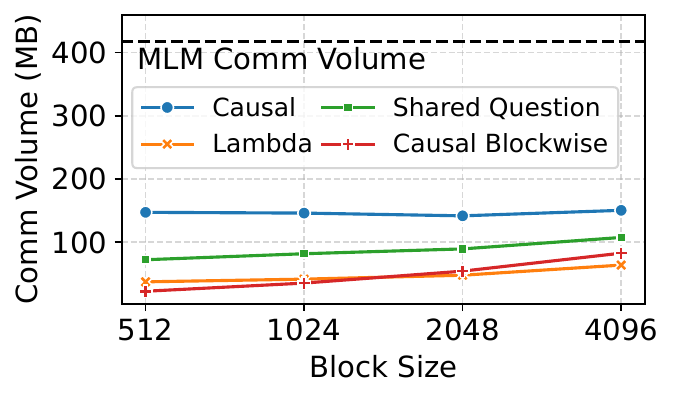}
         \vspace{-6mm}
         \caption{LongAlign.}
         \label{fig:block_size_vs_total_comm_vol_longalign}
     \end{subfigure}
     \hfill
     \begin{subfigure}[t]{0.49\linewidth}
         \centering
         \includegraphics[width=\textwidth]{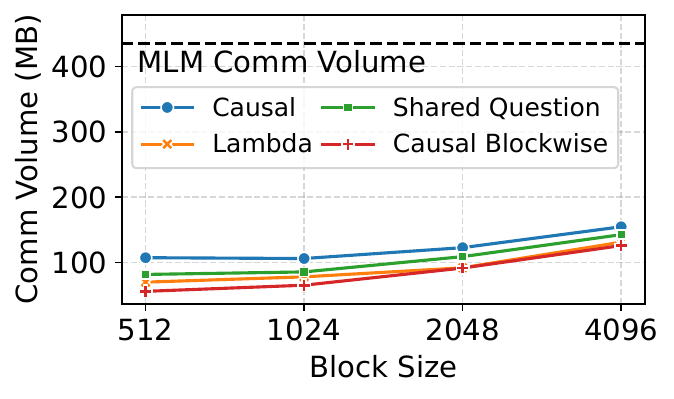}
         \vspace{-6mm}
         \caption{LongDataCollections.}
         \label{fig:block_size_vs_total_comm_vol_ldc}
     \end{subfigure}
     \vspace{-4mm}
    \caption{Total inter-node communication volume v.s. block size.}
    \label{fig:block_size_comm_analysis}
    \vspace{-2mm}
\end{figure}

\begin{figure}
     \centering
     \begin{subfigure}[t]{0.49\linewidth}
         \centering
         \includegraphics[width=\textwidth]{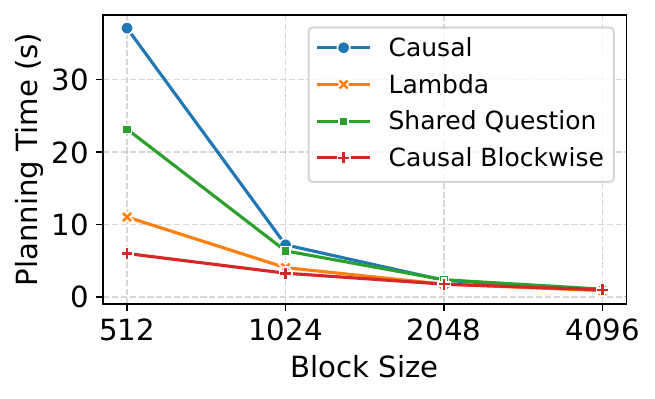}
         \vspace{-6mm}
         \caption{LongAlign.}
         \label{fig:planning_time_longalign}
     \end{subfigure}
     \hfill
     \begin{subfigure}[t]{0.49\linewidth}
         \centering
         \includegraphics[width=\textwidth]{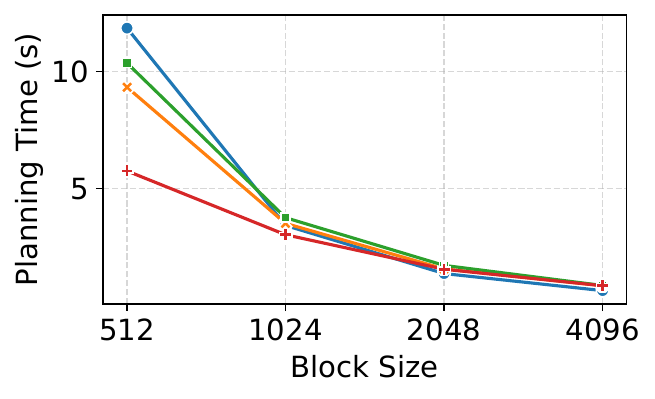}
         \vspace{-6mm}
         \caption{LongDataCollections.}
         \label{fig:planning_time_ldc}
     \end{subfigure}
     \vspace{-4mm}
    \caption{Impact of block size on planning time.}
    \label{fig:block_size_planning_time_analysis}
    \vspace{-2mm}
\end{figure}

\begin{figure}
     \centering
     \begin{subfigure}[t]{0.49\linewidth}
         \centering
         \includegraphics[width=\textwidth]{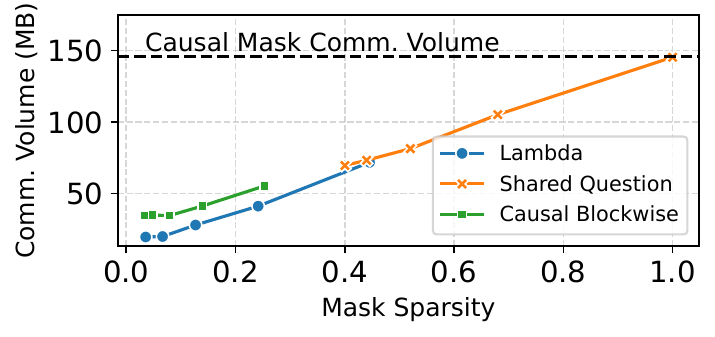}
         \vspace{-6mm}
         \caption{LongAlign.}
         \label{fig:mask_sparsity_longalign}
     \end{subfigure}
     \hfill
     \begin{subfigure}[t]{0.49\linewidth}
         \centering
         \includegraphics[width=\textwidth]{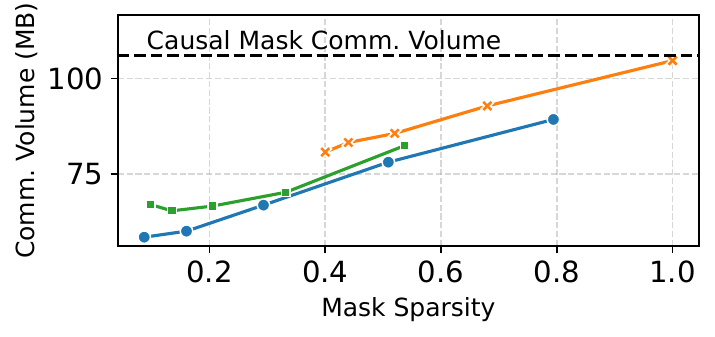}
         \vspace{-6mm}
         \caption{LongDataCollections.}
         \label{fig:mask_sparsity_ldc}
     \end{subfigure}
     \vspace{-4mm}
    \caption{Impact of mask sparsity on communication.}
    \label{fig:mask_sparsity}
    \vspace{-1mm}
\end{figure}

\begin{figure}
 \centering
 \includegraphics[width=0.8\linewidth]{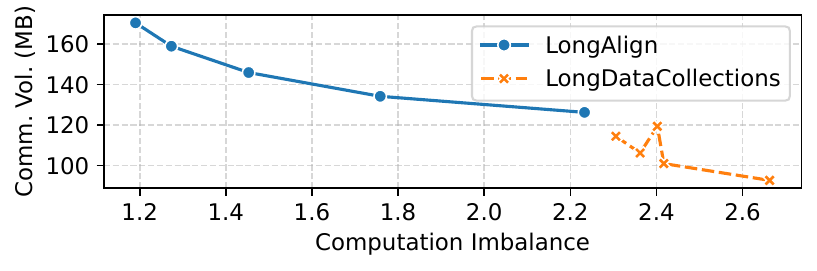}
 \vspace{-5mm}
 \caption{Impact of computation imbalance tolerance on communication.}
 \label{fig:comp_imbalance}
\end{figure}

We further analyze the performance of DCP in detail.
We use the same experimental setups as in the end-to-end experiment (\S\ref{sec:e2e_eval}).

\vspace{1mm}
\noindent\textbf{Communication vs.~block size.}
Fig.~\ref{fig:block_size_comm_analysis} shows the impact of the block size on total inter-instance communication volume and maximum per-device communication volume (including send and receive).
DCP requires much less communication than the MLM baseline (Fig.~\ref{fig:block_size_vs_total_comm_vol_longalign},\ref{fig:block_size_vs_total_comm_vol_ldc}), with communication volume slightly increasing with the block size, since a larger block size (and thus a smaller total block number) provides less placement flexibility.

\vspace{1mm}
\noindent\textbf{Planning time vs.~block size.}
Fig.~\ref{fig:block_size_planning_time_analysis} shows the influence of the block size on planning time (including block generation, hypergraph partitioning and computation/communication scheduling).
As the block size increases, planning time rapidly decreases since the time is directly related to the total number of blocks.
For similar reasons, the planning time is much smaller under sparse attention masks.
When choosing a reasonable block size, the average planning time is less than 10 seconds per training batch/iteration, which can perfectly overlap model execution time (> 1 second per iteration) using our pre-fetching and parallel planning design if planning is parallelized with more than 10 CPU cores (much less than the available number of cores for a typical training server, e.g., 96 as in AWS EC2 \texttt{p4de.24xlarge}).

\vspace{1mm}
\noindent\textbf{Communication vs.~mask sparsity.}
In Fig.~\ref{fig:mask_sparsity}, the mask sparsity is computed as the amount of computation (FLOPS) required for the sparse mask divided by that of the causal mask.
We observe that the required communication volume with DCP grows nearly linearly with mask sparsity, indicating that DCP can exploit the mask sparsity well in eliminating redundant communication.

\vspace{1mm}
\noindent\textbf{Communication vs.~computation imbalance tolerance.}
In Fig.~\ref{fig:comp_imbalance}, the required communication decreases as we increase computation imbalance tolerance ($\epsilon$), indicating a clear trade-off between computation imbalance and required communication. In communication-bound scenarios, $\epsilon$ should be set to a larger value, and vice versa.

\subsection{Precision}

\begin{figure}
% \captionsetup[subfigure]{justification=centering}
     \centering
     \begin{subfigure}[t]{0.49\linewidth}
         \centering
         \includegraphics[width=\textwidth]{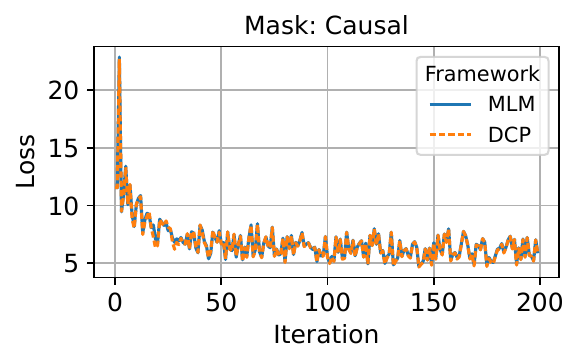}
         \vspace{-6mm}
         \caption{Causal Mask.}
         \label{fig:loss_curve_causal_131072}
     \end{subfigure}
     \hfill
     \begin{subfigure}[t]{0.49\linewidth}
         \centering
         \includegraphics[width=\textwidth]{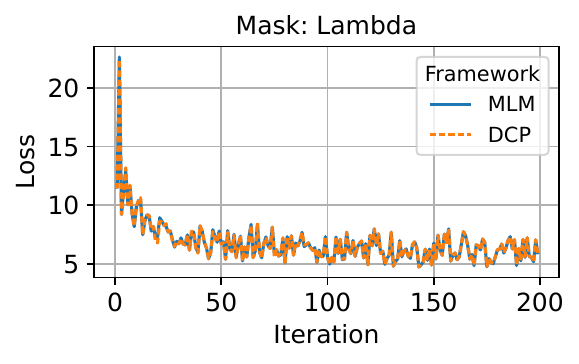}
         \vspace{-6mm}
         \caption{Lambda Mask.}
         \label{fig:loss_curve_lambda_131072}
     \end{subfigure}
     \hfill
     \begin{subfigure}[t]{0.49\linewidth}
         \centering
         \includegraphics[width=\textwidth]{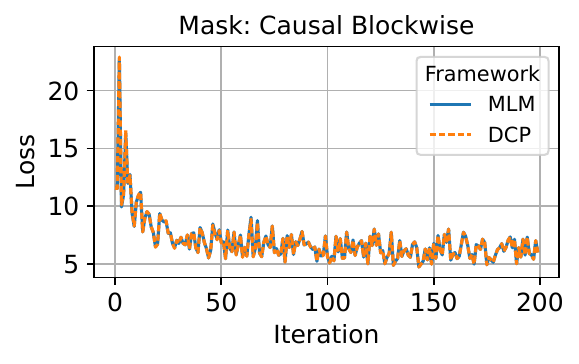}
         \vspace{-6mm}
         \caption{Causal Blockwise Mask.}
         \label{fig:loss_curve_causal_blockwise_131072}
     \end{subfigure}
     \hfill
     \begin{subfigure}[t]{0.49\linewidth}
         \centering
         \includegraphics[width=\textwidth]{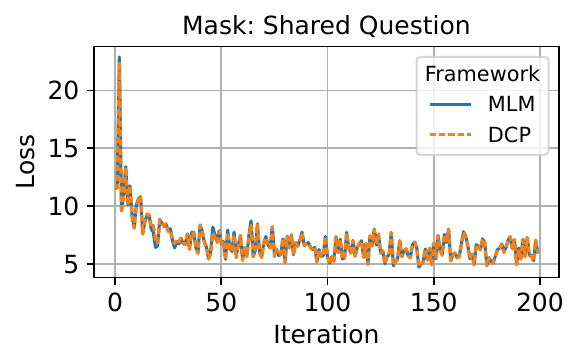}
         \vspace{-6mm}
         \caption{Shared Question Mask.}
         \label{fig:loss_curve_shared_question_131072}
     \end{subfigure}
      \vspace{-2mm}
    \caption{Training loss curves on the LongAlign dataset (max.~sequence length 131072).}
    \label{fig:loss_curve}
    \vspace{-8mm}
\end{figure}

Since DCP does not alter the attention algorithm, we expect it to have no impact on training accuracy.
To verify this, we compare the training loss curves of DCP with those of standard head and ring parallel attention, as implemented in the MLM baseline.
We use the same setup as in the end-to-end experiments.
In Fig.~\ref{fig:loss_curve}, DCP's loss curve matches that of the MLM baseline, only with small deviations due to different kernel implementations and attention/reduction computation orders.

\subsection{Speed-up decomposition}
\label{sec:decomposition}
\begin{figure}
     \centering
     \includegraphics[width=0.8\linewidth]{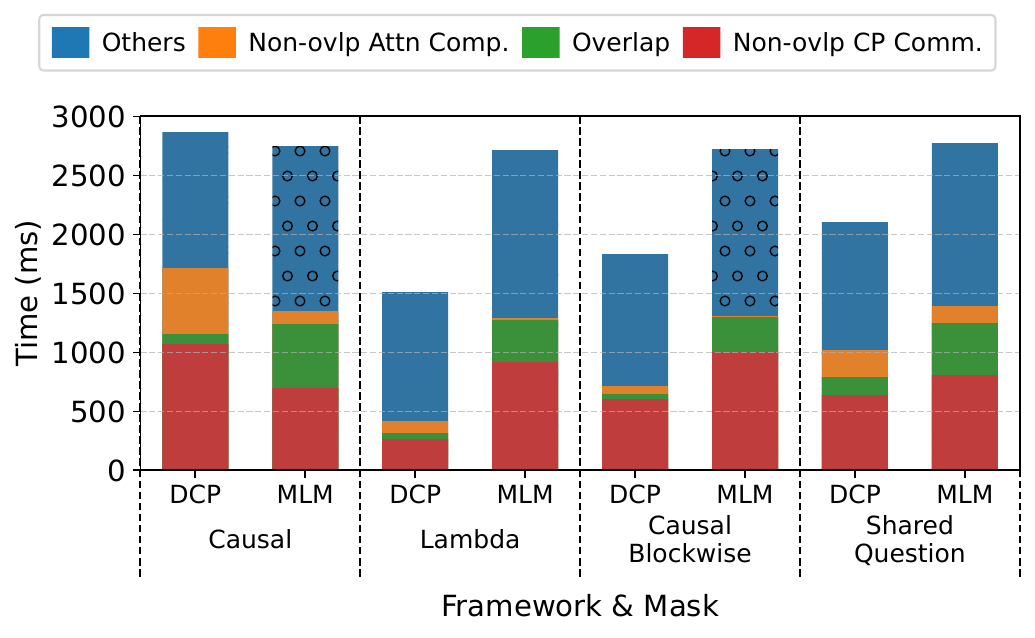}
    \vspace{-4mm}
    \caption{Decomposition of end-to-end iteration time (LongAlign dataset, max sequence length 131072).}
    \label{fig:iter_time_decomp_comparison}
    \vspace{-6mm}
\end{figure}

Fig.~\ref{fig:iter_time_decomp_comparison} illustrates the decomposition of end-to-end iteration time, computed from NVIDIA Nsight Systems\cite{nvidia2025nsightsystems} traces collected during iterations 50–55, with the same setup as Fig.\ref{fig:e2e_longalign_131072}.
For sparse masks, DCP significantly reduces the total communication time (i.e., Non-overlap CP communication + Overlap).
The total attention computation time is also slightly reduced, especially for highly sparse lambda and causal-blockwise masks.
Most of this reduction comes from the attention backward kernel:
MLM uses a fixed number of backward steps, each incurring a certain overhead (i.e., reading local Q and KV from memory, writing gradients back, and performing additional gradient reduction operations across different blocks).
This overhead is more significant in the backward pass.
We observe that DCP's scheduler tends to concentrate all backward computation into one or two divisions (see Section~\ref{sec:comp_comm_scheduling}) when using sparse masks, thus reducing the overall overhead and computation time.
In the causal mask case, where DCP slightly underperforms MLM, the total communication time is still reduced; however, the overlap between computation and communication decreases noticeably. 
We attribute this to limitations in the scheduling algorithm and believe further research could improve its performance.

%% file: sections/8_related_work.tex
\section{Discussion and Related Works}

\noindent
\textbf{Scaling to larger models/clusters.} 
As model or cluster size increases, DCP’s planning overhead is expected to scale sub-linearly for a given input: 1) planning is independent of model size; 2) graph partitioning primarily depends on the number of input blocks, not cluster size; and 3) greedy scheduling scales linearly with the number of devices.
Thus the planning-to-execution time ratio decreases with system scaling.
Batch size scaling can be managed by adding CPU resources or grouping nodes, applying DCP within groups and traditional DP across groups. 
Performance-wise, DCP’s optimizations are unaffected by hidden size or layer count, as layers share the same structure.
Expanding context parallelism to more devices increases communication overhead, underscoring the importance of our communication optimizations.

\noindent
\textbf{Related Works.}
Current context-parallel frameworks like RingAttention~\cite{liu2024ringattention}, USP~\cite{fang2024usp}, LoongTrain~\cite{gu2024loongtrain} and TransformerEngine~\cite{nvidia2024transformerengine} employ a static parallelization configuration that does not adapt to input dynamism.
Hierarchical Balance Packing\cite{yao2025hierarchicalbalancepackingefficient} and WLB-LLM~\cite{wang2025wlbllm} mitigate imbalanced computation within data and pipeline parallelism caused by input sequence length variance via optimizing packing algorithms.
WLB-LLM further discusses the imbalanced computation in context parallelism caused by applying partitioning directly on packed inputs.
In this work, our discussion assumes context parallelism partitioning is performed at the sequence (i.e., each sample/document in the batch) level, which does not exhibit such imbalance.
ByteScale~\cite{ge2025bytescale} and FlexSP~\cite{wang25flexsp} allow different sequences to be parallelized differently (DP v.s. CP/SP) to minimize communication, which is similar to our objective.
However, they do not model fine-grained token dependencies and thus do not support various sparse or structured attention patterns.

\section{Conclusion}
We propose DCP, a context parallel training framework optimized for input dynamism. We enable fine-grained parallelism control for each input sequence, optimize data and computation placement via hypergraph partitioning and design efficient planner and executor modules to minimize planning and execution overhead.
Extensive evaluation shows that DCP achieves up to 0.94x\textasciitilde{}1.16x end-to-end speed-up training a model with causal mask, and 1.00x\textasciitilde{}1.46x with sparse masks.